\documentclass[a4paper, UKenglish, cleveref, autoref, thm-restate, nolineno]{socg-lipics-v2021}
\pdfoutput=1 

\hideLIPIcs  

\title{Hyperbolic Sphericity} 

\author{Thomas Bläsius}{Karlsruhe Institute of Technology, Germany}{thomas.blaesius@kit.edu}{https://orcid.org/0000-0003-2450-744X}{}
\author{Lennart Großkreutz}{Karlsruhe Institute of Technology, Germany}{lennart.grosskreutz@student.kit.edu}{}{}
\author{Jean-Pierre {von der Heydt}}{Karlsruhe Institute of Technology, Germany}{jean-pierre.heydt@kit.edu}{https://orcid.org/0009-0000-3852-350X}{This work was supported by funding from the pilot program Core--Informatics of the Helmholtz Association (HGF).}

\authorrunning{T. Bläsius, L. Großkreutz and J.-P. von der Heydt} 

\Copyright{Thomas Bläsius, Lennart Großkreutz and Jean-Pierre {von der Heydt}} 



\category{} 

\relatedversion{} 

\EventEditors{John Q. Open and Joan R. Access}
\EventNoEds{2}
\EventLongTitle{42nd Conference on Very Important Topics (CVIT 2016)}
\EventShortTitle{CVIT 2016}
\EventAcronym{CVIT}
\EventYear{2016}
\EventDate{December 24--27, 2016}
\EventLocation{Little Whinging, United Kingdom}
\EventLogo{}
\SeriesVolume{42}
\ArticleNo{23}
 
\usepackage{mathtools}
\usepackage{todonotes} 
\usepackage{cite}

\definecolor{pastel-green}{rgb}{0.47, 0.87, 0.47}

\DeclareMathOperator\acos{acos}
\DeclareMathOperator\asin{asin}
\DeclareMathOperator\acosh{acosh}

\DeclareMathOperator\acot{acot}

\DeclareMathOperator\distanceE{\mathit{d_E}}
\DeclareMathOperator\distanceH{\mathit{d_H}}

\newcommand\ESph[1]{\ensuremath{\operatorname{ESph}  (#1)}}
\newcommand\HSphs[1]{\ensuremath{\operatorname{HSph} (#1)}}
\newcommand\HSph[2]{\ensuremath{\operatorname{HSph}_{#2}  (#1)}}
\newcommand\Cub[1]{\ensuremath{\operatorname{Cub}(#1)}}

\newcommand\upperHalfspaceModel[1]{\ensuremath{ \mathbb{U}^{#1}  }}

\begin{document}

\maketitle

\begin{abstract}
	The \emph{sphericity} of a graph is the minimum dimension $d$ such that the graph has an intersection representation of $d$-dimensional balls of equal radius.
	While sphericity has been studied in Euclidean space, we initiate the study of \emph{hyperbolic sphericity}.
	The hyperbolic sphericity of a graph can be significantly smaller than its Euclidean counterpart, but, 
	contrary to the Euclidean setting, depends strongly on the radius of the balls.

	We show that, if the radius of the balls can be chosen depending on the graph, the hyperbolic sphericity is upper bounded by the Euclidean sphericity.
	This extends a previous result for $2$-dimensional hyperbolic space, i.e., uniform disk graphs, to arbitrary dimensions.
	Moreover, our proof is significantly simpler. 
	If we fix the radius, i.e., do not make it dependent on the graph, we show that hyperbolic sphericity can be larger than Euclidean sphericity, but by at most~$1$.
	Additionally, we study how hyperbolic sphericity changes with the ball radius.
	We show that choosing a larger radius can substantially decrease the sphericity while increasing it by at most~$1$.
  We also provide a construction of a graph where the sphericity oscillates between different values as the radius increases.

	Besides being theoretically interesting, we note that these results are relevant for graph embeddings in machine learning, where one is interested in low-dimensional numeric representations of symbolic data like graphs.
\end{abstract}

\section{Introduction}\label{sec:introduction}

Graph embeddings map each vertex of a graph to a position in a geometric space,
providing a geometric representation of the graph's structure.
Ideally, neighboring vertices are positioned close together, while non-neighboring vertices are farther apart.
An obvious application of embeddings is graph visualization~\cite{battistaGrapDrawing, kaufmannDrawingGraphs}.
Moreover, embeddings provide a numeric representation of symbolic data, which is essential for machine learning techniques such as link prediction, node classification, or community detection~\cite{node2vec, deepwalk, verse}.

In both applications, having a \emph{low-dimensional} embedding that \emph{accurately} represents the graph is key.
For visualization purposes, it is clear that fewer dimensions (ideally $2$ or $3$) are advantageous~\cite{jacomyForceAtlas2}. 
Moreover, while inaccurate representations can be visually pleasing, they can also be misleading~\cite{bachEdgeBundling}.
For machine learning, low-dimensional representations are less prone to overfitting and are important for computational efficiency. 
Accuracy is clearly necessary for good performance. 

One way to formalize the required dimensionality for accurately representing a graph is via the concept of sphericity, introduced by Maehara~\cite{Mae-Main}.
The \emph{sphericity} of a graph $G$ is the smallest dimension $d$ such that $G$ can be represented as an intersection graph of $d$-dimensional Euclidean unit balls.
Equivalently, each vertex can be assigned a position such that two vertices are connected if and only if their distance is at most~$2$.

From an algorithmic perspective, determining the sphericity of a given graph is NP-hard and even $\exists\mathbb{R}$-complete~\cite{HeinzUnitDiskRecocNPHard, SchaeferDiskRecocERHard}.
However, multiple structural results provide upper bounds on the sphericity.
For example, every graph with $n$ vertices has sphericity at most $n-\omega$, where $\omega$ is the size of the largest clique~\cite{Mae-Main}. 
Using the Johnson-Lindenstrauss lemma~\cite{JohnsonAndLindenstrauss},
Frankl and Maehara~\cite{Mae-Frankl-JL-Lemma} show that the sphericity for graphs with maximum degree~$\Delta$ is bounded by $O(\Delta^2\log n)$.

Some lower bounds are known as well.
For instance, the complete bipartite graph with $2k$ vertices has a sphericity of at least $k$~\cite{Mae-Knn}.
Furthermore, even structurally simple graphs such as stars exhibit a high sphericity of at least $\Omega(\log n)$ in Euclidean space~\cite{Mae-Knn}.
The reason for this is, that there is insufficient space close to the center of the star to host all leaves without them being close to each other.

This limitation highlights a fundamental challenge of Euclidean embeddings to represent graphs with highly skewed degree distributions.
Such distributions commonly appear in real-world graphs,
like social networks, whose degree distributions often follow a power law~\cite{voitalovScaleFree}.
Interestingly, unlike in Euclidean geometry, skewed degree distributions naturally emerge in hyperbolic geometry~\cite{krioukovComplexNetworks}.
The crucial difference is that hyperbolic space expands exponentially,
i.e., the volume of a ball with radius~$r$ grows exponentially ($2^{\Theta(r)}$) while it grows polynomially ($\Theta(r^d)$) in $d$-dimensional Euclidean space.
This provides more space for separating non-adjacent vertices while keeping them close to a common center, making it possible, for example, to represent stars, or in fact arbitrary trees, as $2$-dimensional hyperbolic disk graphs of equal radius~\cite{SakarLowDistortionTrees}, see~\cref{fig:tree} for such an embedding.

Although hyperbolic disk graphs have been studied before~\cite{sandorHyperbolicBall, blaesiusStrongyHyperbolic, blaesiusStructureAndIndependence}, higher-dimensional hyperbolic ball graphs have not been considered much, particularly from a graph theoretic perspective.
To build a better understanding of how well graphs can be represented in different geometric spaces,
we introduce the notion of \emph{hyperbolic sphericity}, which is defined as the smallest dimension $d$ in which a graph can be represented as an intersection graph of equally sized balls in $d$-dimensional hyperbolic space.

A key difference between Euclidean and hyperbolic sphericity is the dependence on the radius of the balls.
In Euclidean space, the intersection representation can be scaled uniformly, making the concrete choice of the radius irrelevant.
In hyperbolic space, however, this scaling invariance does not hold, and the radius strongly influences which structures can be represented.
A useful intuition for this is that small patches of hyperbolic space behave similarly to Euclidean space,
while in larger patches, the exponential expansion becomes more significant.
As an example, the $k\times k$ grid graph on $k^2$ vertices is a hyperbolic uniform disk graph for radii less than $1/k^6$ but not for large radii~\cite{blaesiusStructureAndIndependence}.
On the other hand, stars on $n$ vertices require a radius of at least $\log n$ to be hyperbolic uniform disk graphs~\cite{blaesiusStructureAndIndependence}.
These observations raise two natural questions:

\begin{enumerate}
	\item How does Euclidean sphericity compare to hyperbolic sphericity, depending on the chosen ball radius?
	\item What is the optimal radius for representing a given graph in hyperbolic geometry?
\end{enumerate}


\subparagraph*{Contribution.}
Towards the first question, we investigate the behavior of hyperbolic sphericity for extremely small and large radii.
Because small patches behave similarly to Euclidean space, 
it seems natural that hyperbolic sphericity can be at most as large as Euclidean sphericity for sufficiently small radii.
However, a formal proof still requires some work.
For $2$-dimensional hyperbolic space this was already shown in~\cite{blaesiusStrongyHyperbolic}, and we generalize this result to arbitrary dimensions and provide a significantly simpler proof.
At the other end of the spectrum, some graphs that are Euclidean unit disk graphs cease to be hyperbolic uniform disk graphs when the radius becomes large.
We prove that the resulting increase in required dimensions is bounded:
for every graph and every radius, the hyperbolic sphericity using an arbitrary radius exceeds the Euclidean sphericity by at most~$1$.

Regarding the second question,
we prove that larger radii never increase the hyperbolic sphericity by more than~$1$ compared to the optimal radius.
This provides a strong robustness guarantee for using hyperbolic embeddings.
Instead of having to search for the optimal radius, choosing a sufficiently large radius always yields a near-optimal embedding dimension.

Beyond that, we provide a construction of a graph where the hyperbolic sphericity can increase and decrease multiple times as the radius increases.
More precisely, this graph has a $2$-dimensional representation for small and large radii, but requires at least $3$~dimensions for radii in between. 
This shows that hyperbolic sphericity can have complex behavior depending on the radius and is neither monotonically decreasing nor increasing.
For our analysis of this graph, we transfer a result about $k$-gons with bounded side lengths from Euclidean geometry to hyperbolic geometry, showing that the area is maximized when it is regular with maximum side length.
Furthermore, we investigate the angles of a hyperbolic quadrilateral with constrained side lengths,
identifying a surprising behavior where the quadrilateral possesses two locally minimal configurations.
Either all four sides have the same length, or three sides and one diagonal have the same length.

\subsection{Related Work}

Our work lies at the intersection of geometric graph representations, hyperbolic geometry, and graph embeddings.
We summarize relevant prior work from these areas below.

\subparagraph*{Hyperbolic Ball Graphs.}

To the best of our knowledge, intersection graphs of hyperbolic balls in higher dimensions ($d>2$) have only been studied by Kisfaludi-Bak~\cite{sandorHyperbolicBall}.
The work establishes bounds on the tree-width (depending on $d$), from which efficient algorithms can be derived. 
The runtime for $d+1$-dimensional hyperbolic intersection graphs behaves the same as the runtime for $d$-dimensional Euclidean graphs.
This is consistent with our results, as every $d$-dimensional Euclidean ball graph is also a $d+1$-dimensional hyperbolic ball graph.

In $2$ dimensions, further studies have examined structural and algorithmic properties of hyperbolic disk graphs.
One study analyzes the formation of cliques and the performance of greedy routing in hyperbolic disk graphs~\cite{blaesiusStrongyHyperbolic}.
These results require that the connection radius equals the ground space radius,
which allows for hierarchical structures but prevents grid-like structures.
Another study allows for different radii and shows that large radii lead to separators that can be covered with few cliques and an efficient independent set algorithm~\cite{blaesiusStructureAndIndependence}.

\subparagraph*{Hyperbolic Random Graphs.}

In a setting with random positions and $2$ dimensions, hyperbolic random graphs (HRGs) were first introduced in~\cite{krioukovComplexNetworks}.
In this model, points are sampled in a disk of fixed radius $R$ and two points are connected if the distance between them is less than $R$.
These graphs are of particular interest because they exhibit key features of real-world networks,
including a heterogeneous degree distribution that follows a power law, as well as a high clustering coefficient~\cite{gugelmannHyperbolicDegreeSequence}.
Several structural properties of HRGs are well understood, e.g., HRGs have a giant connected component~\cite{michelLargestComponentHyperbolic, marcosSecondLargestComponent}, logarithmic diameter~\cite{FriedrichDimeterHyperbolicRandomGraphs, KiwiBoundDiameter, muellerDiameterKPKVB}, large cliques~\cite{BlaesiusCliquesHyperbolicRandomGraphs, BaguleyCliqueNumberDegeneracy}, and comparatively small treewidth~\cite{BlaesiusHyperbolicSeparatorsAndTreewidth}.
Furthermore, several classical problems can be solved more efficiently on HRGs, such as vertex cover~\cite{blaesiusVertexCover, blaesiusVertexCoverApproximation}, computing the clique size~\cite{EnujinComputingMaximumCliques, BlaesiusCliquesHyperbolicRandomGraphs}, and shortest paths~\cite{blaesiusShortestPathHyperbolic}.

\subparagraph*{Hyperbolic Graph Drawing.}

Hyperbolic geometry has also been applied to graph visualization, especially for networks with hierarchical structure or skewed degree distributions that align well with the exponential growth of hyperbolic space.
The Poincaré disk, for example, offers a ``fish-eye'' view that allows to focus on specific parts of a graph while pushing the rest of the representation to the boundary of the disk~\cite{lampingHyperbolicBrowser}.
Extensions to $3$-dimensional hyperbolic drawings have also been explored~\cite{MunzerHyperbolic3DDrawing2, MunzerHyperbolic3DDrawing4, MunzerHyperbolic3DDrawing1, MunzerHyperbolic3DDrawing3}.
Several methods can be considered to compute such embeddings:
Log-likelihood embedders attempt to find positions that maximize the probability that the embedded graph can be drawn from a random graph model~\cite{blaesiusEmbeddingHyperbolicPlane, fragkiskosNetowrInferenceCommonNeighbors, fragkiskosNetworkMapping}.
Hyperbolic spring embedders apply forces to the vertices that attract neighbors and repel non-neighbors~\cite{blaesiusForceDirected, stephenNonEuclideanEmbedder}.
Other approaches derive the embedding from a spanning tree, with applications in greedy routing~\cite{blaesiusHyperbolicGreedyRouting, eppsteinSuccinctGreedyRouting, kleinbergGeographicRouting}.
All these visualization methods share the tendency to place neighboring vertices close to each other and non-neighboring vertices far apart.
Our results support these approaches, demonstrating that hyperbolic geometry is especially effective for low-dimensional representations of graphs.

At first sight, this appears to contradict Eppstein's conclusion that for certain drawings the hyperbolic plane offers no advantages over Euclidean graph drawings~\cite{eppsteinLimitsHyperbolicGraphDrawing}.
This however, is resolved at second glance, as Eppstein's work deals specifically with distances and angles in planar straight-line drawings, while our work is based on intersection representations.

\subparagraph*{Other Geometric Graph Dimensions.}
Similar to sphericity, the \emph{cubicity} or \emph{boxicity} of a graph $G$ asks for the minimal dimension $d$ such that $G$ can be represented as an intersection graph of $d$-dimensional axis-parallel unit cubes or boxes, respectively.
A theorem by Roberts~\cite{robertsBoxicityCubicity} shows that the cubicity of a complete multipartite graph is given by
$\Cub{K_{n_1,\dots,n_q}} = \sum_{i=1}^q\lceil\log(n_i)\rceil $.
As a consequence, the cubicity of a star and even the complete bipartite graph grows logarithmically with the number of vertices.
For boxicity, the situation differs: since boxes can be stretched into long shapes, stars or complete bipartite graphs can be represented with only $1$ or $2$ dimensions. 
Roberts also gave an example of a graph, where the boxicity is significantly larger than sphericity.
He proved that removing a perfect matching from the complete graph on~$2n$ vertices yields a graph whose boxicity grows linearly~\cite{robertsBoxicityCubicity}.
It is easy to see that this graph has sphericity~$2$.
For this, place the centers of all disks on a common circle such that non-adjacent vertices are placed opposite each other.
In general, boxicity and sphericity are incomparable, and neither can be bounded by the other.
We note that both concepts are difficult to transfer to hyperbolic space, as there is no concept of axis-parallel boxes.

Another related notion of dimensionality is the \emph{dot-product dimension}.
It is defined as the smallest dimension $d$ such that the vertices of a graph can be represented as points in $\mathbb{R}^d$, such that two vertices are adjacent if and only if their dot product is larger than~$1$.
If all vectors are normalized, the dot-product dimension behaves similar to sphericity, but for vectors of different length expressiveness increases.
Stars can be represented with a constant number of dimensions while
the $K_{n,n}$ still requires a linear number of dimensions, even with vectors of different lengths~\cite{ReitermanKnnDotProductDim}.
It is worth noting that $\mathbb{R}^d$ together with the dot-product is not a metric space.
As a result, counterintuitive phenomena can occur, such as two vertices having the same representation without being adjacent because their length is too small.

\subsection{Outline}

The remainder of this paper is structured as follows.
In \cref{sec:preliminaries}, we introduce basics of hyperbolic geometry and sphericity.
We also provide some simple lemmas regarding hyperbolic geometry.
In \cref{sec:relation}, we prove our results comparing hyperbolic and Euclidean sphericity for small and large radii.
Finally, in \cref{sec:special_graphs} we give the construction of a graph with oscillating hyperbolic sphericity.
To enable this construction, we prove properties of $k$-gons with maximum area in \cref{sec:area-maximum-k} and use \cref{sec:nasty-quadrilateral} to analyze the angular behavior of hyperbolic quadrilaterals. 
We end with several open questions in \cref{sec:open_problems}.
	
\section{Preliminaries}\label{sec:preliminaries}

\begin{figure}
	\centering
	\centering
	\begin{minipage}[t]{0.63\textwidth}
		\centering
		\includegraphics[page=1]{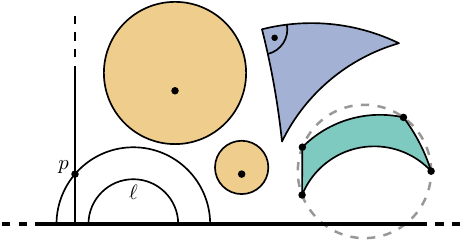}
    \subcaption{Upper half-space}\label{fig:halfspace_example1}
    \end{minipage}
	\hfill
	\begin{minipage}[t]{0.36\textwidth}
		\centering
		\includegraphics[page=2]{graphics/halfspace_example}
        \subcaption{Poincaré disk}\label{fig:halfspace_example2}
    \end{minipage}
	\caption{
		Depiction of $2$-dimensional hyperbolic geometry in the upper half-space  and Poincaré disk model.
    In the half-space model (a) the $x$-axis (thick line) represents the boundary. 
		There are $3$ hyperbolic geodesics, one of which is an open vertical ray and two are open semicircles orthogonal to the boundary.
		For the given geodesic $\ell$ and the point $p$ not on it, there is more than one geodesic containing $p$ that does not intersect $\ell$.
    In the Poincaré disk model (b) the boundary is represented by the outer circle (thick line). 
    The geodesics are open circular arc segments or a line through the center of the disk.
		Both models contain two circles (orange) with an equal hyperbolic radius, the circle whose center lies closer to the boundary appears smaller in the Euclidean drawing.
		The triangle (blue) has a right angle.
		Note that the four points of the quadrilateral do not lie on a common hyperbolic circle but on a so called \emph{hypercycle} (dashed circle).
}\label{fig:halfspace_example}
\end{figure} 

Various equivalent models for hyperbolic space exist, each realized as a subset of $\mathbb{R}^d$ together with a suitable metric.
In this paper, we primarily use the upper half-space model, as it is most convenient for our proofs and has an intuitive geometric interpretation.
Some figures also use the Poincaré disk model, which we explain briefly later.
Let the \emph{$d$-dimensional upper half-space} $\upperHalfspaceModel{d}$ be the set of all points $p=(p_1,\dots,p_d)$ where the last coordinate is greater than zero, 
i.e., $\upperHalfspaceModel{d} = \{ (p_1, \dots, p_d) \in \mathbb{R}^d \mid p_d > 0 \}$.
Two points in the model $p,q\in \upperHalfspaceModel{d}$ can also be interpreted as points in Euclidean space, 
and we can assign both a \emph{Euclidean distance} $\distanceE(p, q)$ and a \emph{hyperbolic distance} $\distanceH(p, q)$ between them.
The Euclidean distance is given by the identity $\distanceE(p,q)^2=\sum_{i=1}^{d} (p_i-q_i)^2$ 
and the hyperbolic distance is given by $\cosh(\distanceH(p, q)) = 1 + \distanceE(p, q)^2 /(2 p_d q_d)$, where $\cosh$ is the hyperbolic cosine.
Note that hyperbolic distances tend to infinity as one of the last coordinates $p_d,q_d$ approaches~$0$.

For some intuition, consider the $2$-dimensional drawing of $\upperHalfspaceModel{2}$ in~\cref{fig:halfspace_example1}.
Here, hyperbolic geodesics, i.e., the shortest curves between points, and circles have a particularly simple representation.
Hyperbolic geodesics are either represented as open rays or open Euclidean semicircles that meet the $x$-axis at a right angle.
Note that points on the $x$-axis are not part of the hyperbolic space.
Moreover, the upper half-space model is \emph{conformal}, meaning that the Euclidean angle between two curves in the representation corresponds exactly to the hyperbolic angle.
Hyperbolic circles correspond to Euclidean circles and vice versa, whereby the Euclidean circle lies above the $x$-axis and has a different center and radius from the hyperbolic circle.
More specifically, a hyperbolic circle with center $(p_1,p_2)$ and radius $s$ corresponds to a Euclidean circle with center $(p_1, p_2\cosh(s))$ and radius $r=p_2 \sinh(s)$~\cite[p.~128, 132]{Ratcliffe}.
We usually use $s$ to denote hyperbolic radii and $r$ to denote Euclidean radii.
Another distinct property of hyperbolic space is the relationship between the angles and the area of a triangle:
the sum of internal angles $S$ of a triangle is always strictly less than $\pi$, and the area of the triangle is given by $\pi - S$.

\begin{figure}
	\centering
	\begin{minipage}[t]{0.53\textwidth}
		\centering
		\includegraphics[]{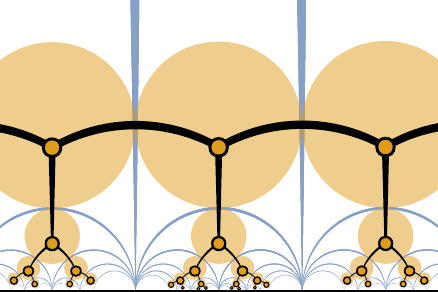}
        \subcaption{Upper half-space}\label{fig:tree1}
    \end{minipage}
	\qquad
	\begin{minipage}[t]{0.37\textwidth}
		\centering
		\includegraphics[]{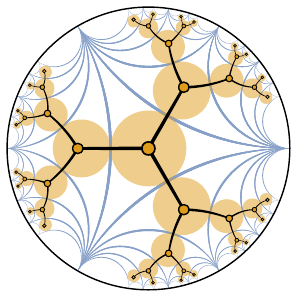} 
        \subcaption{Poincaré disk}\label{fig:tree2}
    \end{minipage} 
	\caption{A $2$-dimensional HUBG representation of a binary tree. Figure (a) uses the upper half-space model and (b) uses the Poincaré disk model. The blue lines separating the circles represent a $\{ 3,\infty \}$-tiling, which allows for a recursive construction of the representation.}\label{fig:tree}
\end{figure}

In addition to the half-space model, we use the \emph{Poincaré disk model} in some of our illustrations, as it is sometimes visually clearer.
The Poincaré disk model consists of all points within an open disk with radius~$1$, where the boundary is not part of the space and represents points at infinity.
This is particularly useful for visualization, as the entire hyperbolic space is represented within a bounded Euclidean region.
As in the half-space model, hyperbolic geodesics are represented as open circular arc segments that meet the boundary at a right angle.
It is also conformal, and hyperbolic circles are represented as Euclidean circles with different Euclidean centers and radii.
See~\cref{fig:halfspace_example2} for a general example and~\cref{fig:tree2} for a drawing of a binary tree.

We briefly define sphericity.
A \emph{graph} $G=(V, E)$ consists of a set of vertices $V$ and a set of undirected edges $E$.
Each edge $\{v_1, v_2\}$ is given by an unordered pair of vertices $v_1,v_2 \in V$. 
For a graph $G$, the \emph{Euclidean sphericity} $\ESph{G}$ is the smallest dimension $d$ 
such that there is a radius $r$ and a mapping $f\colon V\to \mathbb{R}^d$ such that $\distanceE(f(v_1), f(v_2)) \leq 2r$ if and only if $v_1$ and $v_2$ share an edge.
The Euclidean sphericity is independent of the choice of $r$.
In hyperbolic space, the scaling is significant.
The \emph{hyperbolic sphericity with radius $s$} $\HSph{G}{s}$ is the smallest dimension $d$ 
such that there is a mapping $g\colon V\to \upperHalfspaceModel{d}$ with $\distanceH(g(v_1), g(v_2)) \leq 2s$ if and only if $v_1$ and $v_2$ share an edge.
The \emph{hyperbolic sphericity} $\HSph{G}{}$ is the minimum of $\HSph{G}{s}$ over all $s$.
We call the mapping $f$ or $g$ together with the radius a Euclidean or hyperbolic uniform ball graph representation (EUBG or HUBG representation), respectively.
If it is clear from the context, we occasionally identify the vertices of a graph with their positions in a representation. 

\begin{figure}
  \centering
  \begin{minipage}[t]{0.49\textwidth}
    \centering
    \includegraphics[page=1]{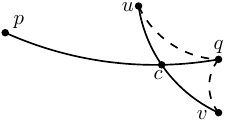}
    \subcaption{Segments from~\cref{lem:planar_drawing}.}\label{fig:triangle_lemmas1}
  \end{minipage}
  \hfill
  \begin{minipage}[t]{0.49\textwidth}
    \centering
    \includegraphics[page=2]{graphics/triangle_lemmas.pdf}
    \subcaption{Triangle from~\cref{lem:extrem_triangle1} and~\cref{lem:extrem_triangle2}.}\label{fig:triangle_lemmas2}
  \end{minipage}
  \caption{Illustration for the short proofs regarding hyperbolic geometry.}\label{fig:triangle_lemmas}
\end{figure}

Before continuing with our main results, we provide some simple lemmas regarding distances and triangles in hyperbolic geometry.
They are used in our construction of the graph in \cref{sec:special_graphs}.
\cref{lem:planar_drawing} allows us to assume that the representation of a graph is planar.
We use~\cref{lem:extrem_triangle1} and~\cref{lem:extrem_triangle2} to bound the angles between neighboring vertices.
Throughout the rest of this section, we use the convention that $A, B, C$ denote the corners of a triangle, $\alpha,\beta,\gamma$ the angles at the corners and $a, b, c$ the lengths of the opposite sides, respectively; see~\cref{fig:triangle_lemmas2}. 

\begin{lemma}\label{lem:planar_drawing}
	Let $\{p,q\}$ and $\{u,v\}$ be two edges in a $2$-dimensional HUBG representation of a graph $G$.
	If the geodesic segment from $p$ to $q$ and from $u$ to $v$ intersect, then the set $\{p,q,u,v\}$ induces at least one triangle in $G$.
\end{lemma}

\begin{proof}
	Let $s$ be the ball radius of the representation and $c$ be the intersection point of the two geodesic segments.
	Without loss of generality, assume that $p$ has a larger distance from $c$ than $q,u$ or $v$; see~\cref{fig:triangle_lemmas1}.
	Using triangle inequality we obtain $\distanceH(q, v)\leq\distanceH(q,c)+\distanceH(c,v)\leq\distanceH(q,c)+\distanceH(c,p)\leq 2s$.
	And in the same way we obtain $\distanceH(q,u)\leq 2s$.
	Therefore, $\{q,u,v\}$ induces a triangle in $G$.
\end{proof}

\begin{lemma}\label{lem:extrem_triangle1}
	For a triangle with $a,b \leq x$, $c > x$ it holds that $\sin(\gamma/2) > \sinh(x/2)/\sinh(x)$.
\end{lemma}

\begin{proof}
	Assume that $\sin(\gamma/2) \leq \sinh(x/2)/\sinh(x)$, we show that $c\leq x$, a contradiction. 
	We claim that $c$ is maximized, when $a,b$ are chosen as large as possible.
	Fix corners $C$ and $B$ and move $A$ along the geodesic through $A$ and $C$, keeping $\gamma$ the same.
	Then the distance $c$ behaves unimodal and is maximized when $b$ is at its boundary values, i.e., $b=0$ or $b=x$.
	If $b=0$, then $A$ and $C$ are equal, and we get $c\leq x$ which is a contradiction.
	Thus, we can assume that $b=x$ and with a symmetric argument assume $a=x$.
  It follows that the triangle $ABC$ is isosceles, and we can split it into two right-angled triangles by the perpendicular bisector of $c$.
	Using our initial assumption $\sin(\gamma/2) \leq \sinh(x/2)/\sinh(x)$ and the sine formula for right-angled triangles $\sin(\gamma/2) = \sinh(c/2)/\sinh(a)$, we get:
	\begin{align*}
		\frac{\sinh(x/2)}{\sinh(x)} \geq \sin(\gamma/2) = \frac{\sinh(c/2)}{\sinh(x)} \implies x \geq c
	\end{align*}
	Again, a contradiction.
\end{proof}

\begin{lemma}\label{lem:extrem_triangle2}
	For a triangle with $a,b > x$, $c \leq x$ it holds that $\sin(\gamma/2)<\sinh(x/2)/\sinh(x)$.
\end{lemma}

\begin{proof}
	The hyperbolic law of cosines states:
	\begin{align*}
		\cos(\gamma) &= \frac{\cosh(a)\cosh(b)-\cosh(c)}{\sinh(a)\sinh(b)}
	\end{align*}
	By applying the half-angle formula $\sin^2(\gamma/2) = (1-\cos(\gamma))/2$ and the addition formula for hyperbolic cosine $\cosh(a-b)=\cosh(a)\cosh(b)-\sinh(a)\sinh(b)$, we obtain:
	\begin{align*}
		\sin^2\left(\frac{\gamma}{2}\right) &= \frac{1}{2}\left(1 - \frac{\cosh(a)\cosh(b)-\cosh(c)}{\sinh(a)\sinh(b)}\right) \\
		&= \frac{\sinh(a)\sinh(b) - \cosh(a)\cosh(b) + \cosh(c)}{2\sinh(a)\sinh(b)} \\
		&= \frac{\cosh(c) - \cosh(a-b)}{2\sinh(a)\sinh(b)}
	\end{align*}
	As $\cosh(x)$ and $\sinh(x)$ are strictly monotone increasing, 
	this shows that $\sin(\gamma/2)$ is maximized by choosing $c$ as large as possible and $a = b$ as small as possible.
	Doing this and using $\cosh(x)-1 = 2\sinh^2(x/2)$, we obtain:
	\begin{align*}
		\sin^2\left(\frac{\gamma}{2}\right) < \frac{\cosh(x) - 1}{2\sinh^2(x)} = \frac{\sinh^2(x/2)}{\sinh^2(x)}
	\end{align*}
	Taking the square root on both sides yields the claim.
\end{proof}



\section{Relation Between Hyperbolic and Euclidean Sphericity}\label{sec:relation}
In this section we provide our theorems comparing Euclidean and hyperbolic sphericity.
First, we show that for every graph $G$ the hyperbolic sphericity is at most the Euclidean sphericity, i.e., $\HSphs{G}\leq \ESph{G}$.
Second, we prove that for arbitrary radii $s$ the hyperbolic sphericity can exceed the Euclidean sphericity by at most 1, $\HSph{G}{s} \leq \ESph{G} + 1$.
Finally, we show that for sufficiently large radii there is an $s_1$ such that for all $s\geq s_1$ $\HSph{G}{s}\leq \HSphs{G}+1$.


\subparagraph*{Small Radius.} We show that any EUBG representation of a graph can be translated into a HUBG representation of the same dimension.
The key observation is that small regions of hyperbolic space behave similarly to Euclidean space.
Thus, shrinking the Euclidean representation and interpreting the resulting coordinates as hyperbolic coordinates only introduces a minor distortion of distances.
This kind of reasoning has already been used for the $2$-dimensional case~\cite{blaesiusStrongyHyperbolic}.
Although the high-level argument is simple, the technical details are somewhat tedious, as one has to make sure that the reduced distortion is not outweighed by the smaller coordinate values.
Here, we generalize the statement for higher dimensions and also substantially simplify the proof.
The simplification comes from the fact that we do not consider distances explicitly, but instead only maintain the order of distances between vertex pairs.
Recall that the hyperbolic distance between two points in $\upperHalfspaceModel{d}$ decreases as they are translated upwards, 
and therefore the observed region of hyperbolic space becomes smaller the farther the points are translated upwards.
Although this is not a uniform scaling operation,
we use the following lemma to show that for any finite point set,
the order of distances no longer changes once they have been shifted far enough upwards. 

\begin{lemma}\label{lem:small-radius-auxilary}
	Let $P\subseteq \upperHalfspaceModel{d}$ be a finite set of points in the upper half-space, and let $P_t$ be the set of points obtained by translating $P$ upwards by $t$.
	There is a $t_0$ such that for all $t\geq t_0$ and all $p,q,u,v \in P_t$, it holds that if $\distanceE(p, q) < \distanceE(u, v)$ then $\distanceH(p, q) < \distanceH(u, v)$.
\end{lemma}
\begin{proof}
	Let $p,q,u,v\in P_t$, we use the notation $p=(p_1,\dots,p_{d-1},p_d+t)$ to refer to the coordinates of a point.
	Recall that the hyperbolic distance between $p$ and $q$ is given by
	\begin{equation*}
		\distanceH(p,q) = \acosh \left( 1 + \frac{ \distanceE(p, q)^2 }{ 2 (p_d+t)(q_d+t) } \right).
	\end{equation*}
	Assume that $d_E(p,q) < d_E(u,v)$.
	We want to show that $d_H(p,q) < d_H(u,v)$ for sufficiently large $t$.
	As $\acosh(1 + x)$ is strictly increasing for $x \in [0,\infty]$, it suffices to show that
	\begin{equation*}
		\frac{ \distanceE(p,q)^2 }{ 2 (p_d+t)(q_d+t) }
		< \frac{ \distanceE(u,v)^2 }{ 2 (u_d+t)(v_d+t) }
		\quad\Leftrightarrow\quad
		\frac{(u_d + t) (v_d + t)}{ (p_d + t) (q_d + t) } <  \frac{ \distanceE(u,v)^2 }{ \distanceE(p,q)^2}.
	\end{equation*}
	For the right-hand side, we have $\distanceE(u,v)^2 / \distanceE(p,q)^2 > 1$ as $\distanceE(p, q) < \distanceE(u, v)$ is independent of $t$.
	Moreover, rearranging the left-hand side yields
	\begin{equation*}
		\frac{(u_d + t) (v_d + t)}{ (p_d + t) (q_d + t) } \le
		\frac{t^2 + u_d t + v_d t + u_d v_d}{ t^2 } =
		1 + u_d / t + v_d/t + u_d v_d/t^2,
	\end{equation*}
	which clearly converges to $1$ for $t \to \infty$.
	Thus, for sufficiently large $t$, $\distanceE(p, q) < \distanceE(u, v)$ implies $\distanceH(p, q) < \distanceH(u, v)$, which concludes the proof.
\end{proof}

Using this lemma, we can show that for small radii the hyperbolic sphericity is always less than or equal to the Euclidean sphericity.
The core idea is as follows:
We use the translation from \cref{lem:small-radius-auxilary}, to transform a $d$-dimensional EUBG representation to a $d$-dimensional hyperbolic one.
Since the ordering between Euclidean and hyperbolic distances is preserved, edges have a smaller distance than non-edges in both representations.

\begin{theorem}\label{thm:small-radius}
  For every graph $G$, there exists some radius $s_0 > 0$ such that for each $s \leq s_0$,
  $\HSph{G}{s} \leq \ESph{G}$. In particular, $\HSphs{G} \leq \ESph{G}$.
\end{theorem}

\begin{proof}
	Consider a $d$-dimensional EUBG representation of $G$ with radius~$\frac{1}{2}$, where the last coordinate for each vertex is positive.
	Let $e_1, \dots, e_k$ be the list of all unordered pairs of vertices from $G$ and $a_1,\dots, a_k$ be the Euclidean distances between the points of the corresponding pair with $k=\binom{n}{2}$.
	We may assume that the distances are sorted in ascending order and by slight perturbation of the positions, we may even assume that they are strictly increasing.
	This way, the first $|E|$ distances correspond to edges and are smaller or equal to~$1$ and the last $k-|E|$ distances correspond to non-edges, being greater than~$1$.

	For a fixed $t$, we construct a $d$-dimensional HUBG representation by translating the last coordinate of each vertex upwards by $t$ and interpreting the coordinates in the upper half-space model.
	Let $b_1(t), \dots, b_k(t)$ be the list of hyperbolic distances of the pairs of translated vertices.
	By \cref{lem:small-radius-auxilary}, there is a $t_0$ such that for all $t \geq t_0$ the distances $b_1(t),\dots,b_k(t)$ are also sorted in ascending order.
	We choose a radius $s(t)$ such that $b_{|E|}(t) < 2 s(t) < b_{|E| + 1}(t)$, e.g., $2s(t) = \frac{1}{2}(b_{|E|}(t) + b_{|E| + 1}(t))$.
	This way, the distance for every edge is less than $2s$, while the distance for every non-edge is larger.
	As $s(t)$ is continuous and approaches $0$ for $t \to \infty$, this is a valid representation for all $s(t) \leq s(t_0)=s_0$. 
\end{proof}

Based on this proof, one could expect that Euclidean and hyperbolic sphericity are equal for sufficiently small radii.
However, this is not always the case.
An example of this is the star with~$6$ leaves, which has a hyperbolic sphericity of~$2$ for all radii but a Euclidean sphericity of~$3$.

\subparagraph*{Arbitrary Radius.}
For large radii, the distortion between Euclidean and hyperbolic geometry increases, and it is reasonable to assume that the representable structures also differ increasingly.
Interestingly, this is only partially true.
Although there are hierarchical structures, such as trees and stars, which have a much greater Euclidean than hyperbolic sphericity for large radii, the reverse does not hold.
In fact, for all graphs the hyperbolic sphericity is at most~$1$ greater than the Euclidean sphericity, even for very large radii.

\begin{figure}
	\centering
	\includegraphics{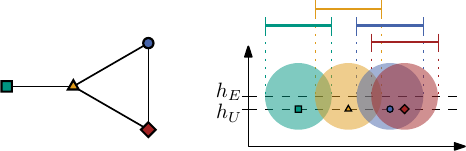}
	\caption{Converting a $1$-dimensional EUBG representation to a $2$-dimensional hyperbolic representation
		by assigning $y_e$ as the second coordinate of each center,
		and then interpreting it as a representation in upper half-space.
		The corresponding hyperbolic centers share $y_h$ as their second coordinate.
	}\label{graph-classes:fig:dimension-increase}
\end{figure}

We first prove a simplified version of this statement via \cref{thm:hyp-sph-better}, as it has a more descriptive geometric interpretation.
Afterwards, we prove a more general version in \cref{thm:large-radius}.
For intuition, consider a $1$-dimensional EUBG as seen in~\Cref{graph-classes:fig:dimension-increase}.
This representation can be embedded in $2$-dimensional Euclidean space, where each vertex has the same last coordinate $h_E$.
The Euclidean balls around the vertices can also be interpreted as hyperbolic balls in the upper half-space model, but with a different center and radius.
As the Euclidean balls all have the same last coordinate and radius, the hyperbolic balls also have the same last coordinate and radius.
Changing $h_E$ in the Euclidean representation influences their hyperbolic radius, and it is possible to obtain any hyperbolic radius this way.

\begin{theorem}\label{thm:hyp-sph-better}
  For every graph $G$ and every radius $s > 0$, $\HSph{G}{s} \leq \ESph{G} + 1$.
\end{theorem}

\begin{proof}
	Consider a graph with Euclidean sphericity $d$ and let $s>0$ be any hyperbolic radius.
	For the HUBG representation we embed the $d$-dimensional EUBG into $(d+1)$-dimensional space and set the last coordinate to $t=2/\sqrt{2\cosh(2s)-2}>0$.
	To see that this is a valid HUBG representation with radius $s$, consider two vertices with coordinates $p,q\in \mathbb{R}^d$.
	Using that the hyperbolic distance between $p$ and $q$ is given by $\distanceH(p, q) = \acosh(1 + \frac{\distanceE(p, q)^2}{2 t^2})$ we see:
	\begin{align*}
		\distanceH(p, q) < 2s
		\iff d_E(p, q) < t\sqrt{2(\cosh(2s) - 1)} = 2.
	\end{align*}
	Thus, two vertices are adjacent in the Euclidean representation if and only if they are adjacent in the hyperbolic representation and any hyperbolic radius can be achieved by this construction.
\end{proof}

Using a result from Maehara~\cite{Mae-Main} that the Euclidean sphericity is at most $|V|-1$,
we get that the hyperbolic sphericity is at most $|V|$ for any radius and in particular always finite.

According to \cref{thm:hyp-sph-better}, choosing a large radius increases the hyperbolic sphericity by a maximum of $1$ compared to the Euclidean sphericity.
We also show, that the hyperbolic sphericity behaves the same way towards itself.
For a graph with a $d$-dimensional HUBG representation, increasing the radius, increases the dimension by at most $1$.

\begin{theorem}\label{thm:large-radius}
	Let $G$ be a graph and $0 < s_1 < s_2$.
	Then $\HSph{G}{s_2} \leq \HSph{G}{s_1} + 1$.
\end{theorem}

\begin{proof}
	For convenience, we write $2s_1 = \acosh(1 + \alpha_1)$ and $2s_2 = \acosh(1 + \alpha_2)$ for some $\alpha_2 > \alpha_1 > 0$.
	For $t=\alpha_2/\alpha_1 - 1 > 0$, define the function $f$, which maps the coordinates of a $d$-dimensional HUBG representation to a $(d+1)$-dimensional representation, as:
	\[
		f\colon \upperHalfspaceModel{d} \to \upperHalfspaceModel{d + 1},
		(p_1, \dots, p_d) \mapsto
		\left(p_1, \dots, p_{d - 1}, \sqrt{\frac{t}{1+t}} p_d, \frac{1}{\sqrt{1+t}}p_d \right).
	\]
	We show that, $d_H(p,q) < 2s_1$ if and only if $d_H(f(p), f(q)) < 2s_2$. 
	First, observe that $f$ does not change the Euclidean distance between points:
	\[
		d_E(f(p), f(q))^2 = \sum_{i=1}^{d-1}(p_i-q_i)^2 + \frac{t}{1+t}(p_d-q_d)^2 + \frac{1}{1+t}(p_d-q_d)^2 = d_E(p, q)^2.
	\]
	Using this identity, we obtain the following equivalence of inequalities for the hyperbolic distance. For any points $p,q\in\upperHalfspaceModel{d}$:
	\begin{align*}%
		     & \distanceH(p, q) < \acosh( 1 + \alpha_1) = 2s_1                             \\
		\iff & \frac{d_E(p, q)^2}{2p_d q_d} < \alpha_1                                        \\
		\iff & \frac{d_E(f(p), f(q))^2}{2f(p)_{d+1}\cdot f(q)_{d+1}} < (1+t)\alpha_1 \\
		\iff & \distanceH(f(p), f(q)) < \acosh(1+(1+t)\alpha_1) = \acosh(1+\alpha_2) = 2s_2.
	\end{align*}%
	%
	By this, it follows that applying $f$ to the original representation yields a $(d + 1)$-dimensional representation with radius $s_2$.
\end{proof}

\section{Graph with Non-Monotonic Hyperbolic Sphericity}\label{sec:special_graphs}

\begin{figure}
	\centering
	\begin{minipage}[t]{0.3\textwidth}
		\centering
		\includegraphics[page=3]{graphics/C_k_l_solution.pdf}
        \subcaption{Schematic representation of $\tilde C$ containing $k=8$ (diamond, green) cycle vertices and $p=33$ (circle, orange) paths evenly distributed among the cycle.}\label{fig:c_k_l_schema}
    \end{minipage}
    \hfill
	\begin{minipage}[t]{0.3\textwidth}
		\centering
		\includegraphics[page=1]{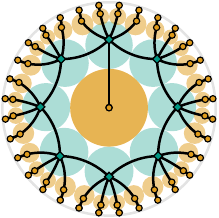}
        \subcaption{Drawing of $\tilde C$ using radius $s_1$. One path is placed inside the cycle. Positions were computed using a force-directed method.}\label{fig:c_k_l_sol1}
    \end{minipage}
    \hfill
	\begin{minipage}[t]{0.3\textwidth} 
		\centering
		\includegraphics[page=2]{graphics/C_k_l_solution.pdf}
        \subcaption{Drawing of $\tilde C$ using radius $s_2$. All paths are placed outside the cycle. Positions were computed using a force-directed method.}\label{fig:c_k_l_sol2}
    \end{minipage}
	\caption{Representations of the graph $\tilde C$ from~\cref{thm:jump_example} for different radii.}\label{fig:c_k_l}
\end{figure}

In this section, we construct a graph for which the set of optimal radii consists of more than one interval.
More specifically, the graph has sphericity $2$ for radii $s_1$ and $s_2$ but sphericity $3$ for radius $s$ in between, i.e., for $s_1 < s < s_2$.
The core idea of our construction is as follows.
Consider the graph $\tilde C$ illustrated in~\cref{fig:c_k_l_schema}.
It consists of a cycle of $k = 8$ vertices together with $p = 33$ paths of length $2$ attached to the cycle vertices.
Assume the radius is sufficiently large, then there is enough space outside the cycle to fit all the paths outside it.
This yields a 2-dimensional HUBG representation with large radius $s_2$; see~\cref{fig:c_k_l_sol2}.
If the radius is sufficiently small, it is possible to fit one path inside the cycle.
If we additionally choose it not too small, there is enough space around the cycle to fit the remaining $p - 1 = 32$ paths around it.
This yields a 2-dimensional HUBG representation for a radius $s_1 < s_2$; see~\cref{fig:c_k_l_sol1}.
In between we can show that there is a radius $s$ with $s_1 < s < s_2$, such that $s$ is too small to fit all $p = 33$ paths outside the cycle, but it is too large to fit one of the paths inside the cycle.

We note that the choices $k = 8$ and $p = 33$ are not completely arbitrary in the sense that there is a dependence between $k$ and $p$.
But, we believe that the construction still works for different values of $k$, yielding different ranges of radii that allow $2$-dimensional HUBG representations.
This might allow us to take the disjoint union of these graphs to obtain a graph where the sphericity oscillates between $2$ and $3$ arbitrarily often for varying radius; see Section~\ref{sec:open_problems} for a discussion.

\begin{figure}
  \centering
  \includegraphics[page=1]{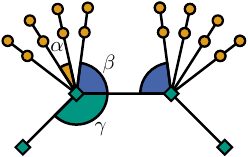}
  \caption{Illustration of angles used in the proof of \cref{thm:jump_example}. 
	All three angle types are shown: type $1$ ($\alpha$, orange) between two path neighbors, type $2$ ($\beta$, blue) between a path neighbor and a cycle neighbor, and type $3$ ($\gamma$, green) between two cycle neighbors.
      }%
      \label{fig:angles} 
\end{figure}

Our proof for the existence of this radius $s$ is heavily based on arguments involving angles.
To this end, we classify the angles at cycle vertices into three different types: angles between two adjacent path vertices (type $1$), between a path and a cycle vertex (type $2$), and between two adjacent cycle vertices (type $3$); see~\cref{fig:angles}.
For the type $3$ angles, we require some general geometric insights into $k$-gons and quadrilaterals.
Specifically, we show in~\cref{sec:area-maximum-k} that a hyperbolic $k$-gon with an upper bound on all side lengths (think of a 2-dimensional HUBG-representation of a $k$-cycle) has maximum area if it forms a regular $k$-gon.
For type $2$ angles, we study the quadrilateral formed by two adjacent cycle vertices and the two path vertices at the type 2 angles.
In~\cref{sec:nasty-quadrilateral} we determine when these angles become minimal.
Finally, we use these more technical insights to show the above claimed existence of $s$ in~\cref{sec:putt-things-togeth}.
For readers that are happy to believe some intuitive geometric assumptions for now, we recommend skipping to~\cref{sec:putt-things-togeth} (although we also find the more technical sections in between insightful).

\subsection{Area Maximum \texorpdfstring{$\mathbf{k}$}{k}-Gons}%
\label{sec:area-maximum-k}

The following lemma shows that among all $k$-gon in which all sides have length at most $x$, the regular $k$-gon in which all sides have length $x$ and all internal angles are equal has maximum area.
It additionally provides a formula for the internal angles we get in this case.

\begin{figure}
  \begin{minipage}[b]{0.32\textwidth}
    \centering
    \includegraphics[page=3]{graphics/kgon_proof.pdf}
    \subcaption{Three consecutive vertices of the $k$-gon.
    For fixed $A$, $C$, and $h$, the area is maximized, when $D$ is the midpoint of $AC$.}\label{fig:kgon1}
  \end{minipage}
  \hfill
  \begin{minipage}[b]{0.32\textwidth}
    \centering
    \includegraphics[page=1]{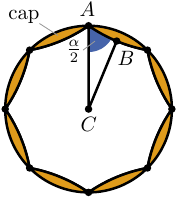}
    \subcaption{Regular $k$-gon and its circumcircle.
    The side of the $k$-gon and the corresponding arc of the circumcircle form a cap (orange).}\label{fig:kgon2}
  \end{minipage}
  \hfill
  \begin{minipage}[b]{0.32\textwidth}
    \centering
    \includegraphics[page=2]{graphics/kgon_proof.pdf}
    \subcaption{A $k$-gon with all equal sides.
    The corresponding caps are the same as in the regular case and have the same area.}\label{fig:kgon3}
  \end{minipage}
  \caption{Illustration of the proof about $k$-gons from~\cref{lem:kgon}.}\label{fig:kgon}
\end{figure} 

\begin{lemma}\label{lem:kgon}
	Given a hyperbolic $k$-gon, with all side lengths $\leq x$ then
	the area of the $k$-gon is maximized, when it is regular and all side lengths are equal to $x$.
	In this case, the angle at any vertex is given by $\alpha = 2\asin\left( \frac{\cos(\pi/k)}{\cosh(x/2)} \right)$.
\end{lemma}
\begin{proof}
  We first show that if the area is maximum, all sides must have the same length.
  We then use this to show that the $k$-gon must be regular.
  From there it is obvious that we get maximum area by using maximum side length.

  To show that all sides have the same length, first note that the $k$-gon is convex. 
  Otherwise, one can clearly increase the area without increasing side length.
  Now let $A$, $B$, and $C$ be three consecutive corners of the $k$-gon.
  We want to show that $|AB| = |BC|$, i.e., the triangle $ABC$ is isosceles.
  For this, let $D$ be the point on $AC$ such that the segment $BD$ is perpendicular to $AC$, i.e., 
  $h = |BD|$ is the height of the triangle; see~\cref{fig:kgon1}.
  This splits the triangle $ABC$ into two right triangles $ABD$ and $BCD$ that share the side $BD$ of length $h$. 
  Note that $ABC$ is isosceles if and only if $D$ is the midpoint of $AC$.
  Thus, we want to show that the area of $ABC$ is maximum if $|AD| = |DC|$.
  Using the standard formula for the area on the two right triangles $ABD$ and $BCD$, we get that the area of $ABC$ is
  \begin{equation*}
    f(|AD|) + f(|DC|) \quad
    \text{with}\quad
    f(x) = 2 \arctan(\tanh( h / 2)\tanh(x / 2)).
  \end{equation*}
  Now it remains to show that this term becomes maximal if $|AD| = |DC| = |AC| / 2$.
  For this, note that $\tanh(h / 2) \cdot \tanh(x / 2)$ is a concave function.
  Moreover, $2\arctan(x)$ is concave and non-decreasing.
  Thus, the concatenation $f(x)$ is also concave~\cite[Section~3.2.4, Equation~3.10]{convex-optimization}.
  Using concavity, we have
  \begin{equation*}
    \frac{f(x) + f(y)}{2} \le f\left( \frac{x + y}{2} \right)
    \quad\Leftrightarrow\quad
    f(x) + f(y) \le 2 f\left( \frac{x + y}{2} \right)
  \end{equation*}
  and thus
  \begin{equation*}
    f(|AD|) + f(|CD|) \le 2 f\left( \frac{|AD| + |DC|}{2} \right)
    = 2 f\left( \frac{|AC|}{2} \right).
  \end{equation*}
  Thus, the area is maximal $|AD| = |DC| = |AC| / 2$.
  
  We now show that the $k$-gon must be regular.
  Essentially, the proof follows the Euclidean argument~\cite[p.~236]{maxima-and-minima}, which states that the area of a $k$-gon with fixed side lengths is maximized, when it is \emph{cyclic}, i.e., all corners lie on a common circle.
  However, we must be careful, as the Euclidean argument requires that the list of side lengths can be realized by a cyclic polygon.
  This is not always the case in hyperbolic geometry; see the quadrilateral in Figure~\ref{fig:halfspace_example}.
  In our case, this is not a problem as all sides have equal lengths and for example the regular $k$-gon is already cyclic.
  
  Since this precondition is satisfied, we can apply the rest of the Euclidean proof.
  In the following, we provide a rough sketch:
  Start by placing all corners on a common circle.
  Consider the \emph{cap} formed by a side of the $k$-gon and the corresponding circular arc segment, see~\cref{fig:kgon2}.
  The area of the circle is equal to the area of the $k$-gon plus the area of all caps.
  Change the shape of $k$-gon while keeping the side lengths fixed and the caps attached to them.
  This changes the area of the $k$-gon but not the area of the caps, as in~\cref{fig:kgon3}.
  Thus, the area of the $k$-gon is maximized, when the area of the $k$-gon plus the area of the caps is maximized.
  For all shapes with fixed perimeter, the area of the shape is maximized by the circle.
  Consequently, the $k$-gon is cyclic and regular.

  To derive the angle formula consider the right-angle triangle formed by the center of the regular $k$-gon $C$, one of its corners $A$ and the midpoint of an adjacent side $B$.
  We obtain $\sin(\alpha/2) = \frac{\cos(\pi/k)}{\cosh(x/2)}$, rearranging yields the formula for $\alpha$.
\end{proof}

\subsection{Quadrilateral with Minimal Angle Sum}%
\label{sec:nasty-quadrilateral}

\begin{figure}[tb]
	\centering
	\begin{minipage}[b]{0.24\textwidth}
		\centering
		\includegraphics[page=1]{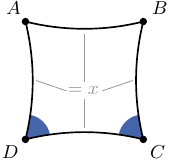}
        \subcaption{First configuration for a quadrilateral with locally minimal angle sum $C+D$.
        All four sides have length equal to $x$, resulting in a regular $4$-gon. 
        The lower bound is derived by~\cref{lem:quadri2}.}\label{fig:quadri_case1}
    \end{minipage}
	\hfill
	\begin{minipage}[b]{0.245\textwidth} 
		\centering
		\includegraphics[page=2]{graphics/quadrilateral_cases.pdf}
        \subcaption{ Second configuration for a quadrilateral with locally minimal angle sum $C+D$.
          Three sides and the diagonal $BD$ have length $x$, $AD$ has length $\leq x$. 
        The lower bound is derived by~\cref{lem:quadri3}.}\label{fig:quadri_case2}
    \end{minipage}
  \hfill
  \begin{minipage}[b]{0.44\textwidth}
		\centering
		\includegraphics[page=1]{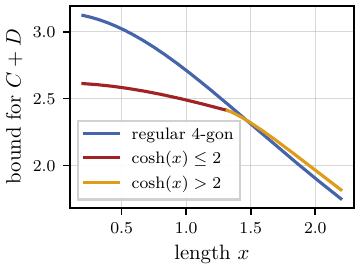}
        \subcaption{Lower bounds on the angle sum $C+D$ as a function of $x$ for the two configurations.
        The second configuration shows a threshold behavior at $\cosh(x) = 2$.}\label{fig:quadri_case3}
    \end{minipage}
	\caption{Two configurations locally minimize the angle sum $C+D$ in the quadrilateral.}\label{fig:quadri_case} 
\end{figure}

In this section, we examine the angular behavior of a hyperbolic quadrilateral with three sides less than or equal to $x$ and one side greater than $x$.
Additionally, the diagonals are also required to have length at least $x$.
We denote the corners with $A$, $B$, $C$, and $D$ such that the side $AB$ is the long side.
For convenience, we call the large side $AB$ \emph{summit}, the two adjacent sides $BC$ and $DA$ \emph{legs}, and the remaining side $CD$ \emph{base}; see Figure~\ref{fig:quadri1-setting}.
To tie this back to the graph we constructed, this is the type of quadrilateral we get when $C$ and $D$ are two adjacent vertices on the cycle, $A$ is a path vertex adjacent to $D$ and $B$ is a path vertex adjacent to $C$; see the type $2$ angles in~\cref{fig:angles}.
The requirement on the lengths reflect the adjacencies.

Our goal is to find out which quadrilateral with the above length constraints minimizes the sum of angles $C + D$ at the base.
Surprisingly, this sum has two local minima, i.e., there are two different configuration in which no vertex can be moved slightly to decrease $C + D$.
This is captured by Lemma~\ref{lem:quadri1}, which shows that in a local minimum, either all four sides have length $x$ or three sides and one diagonal have length $x$; see~\cref{fig:quadri_case}.
For the former case, we provide a bound on $C + D$ in Lemma~\ref{lem:quadri2}.
The case where three sides and one diagonal have length $x$ and the remaining side has length $\le x$ exhibits a rather interesting threshold behavior depending on the $x$; see the red and orange curve in \cref{fig:quadri_case3}.
For $\cosh(x) \leq 2$, the angle sum decreases with shrinking side length, which essentially degenerates the quadrilateral to a triangle.
For $\cosh(x) > 2$ there is a specific positive length smaller $x$ that minimizes the angle sum.
This is formalized in Lemma~\ref{lem:quadri3}.

\begin{lemma}\label{lem:quadri1}
  Let $ABCD$ be a quadrilateral with minimum angle sum $C + D$ such that the sides have lengths $|AB| \ge x$ and $|BC|, |CD|, |DA| \le x$, and the diagonals satisfy $|AC|, |BD| \ge x$.
  If the angles $A$ and $B$ are less than $90^\circ$ then all sides have length $x$.
  If $A \ge 90^\circ$ then all sides except $DA$ have length $x$ and the diagonal $AC$ has length $x$.
\end{lemma}
\begin{proof}
  The arguments for proving the statement usually go as follows.
  We assume that one of the sides does not have the claimed length.
  Based on this, we then show that one of the corners can be slightly moved such that the angle sum $C + D$ is decreased without violating any of the length constraints.
  This contradicts the assumption that this angle sum was minimum among all quadrilaterals satisfying the length constraints.
  As some of these movements are reused for different cases, we organize them in claims, which we prove individually.

  \begin{figure}
    \centering
    \begin{minipage}[t]{0.3\linewidth}
      \centering
      \includegraphics[page=1]{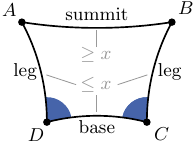}
      \subcaption{General setting in Lemma~\ref{lem:quadri1}.}%
      \label{fig:quadri1-setting}
    \end{minipage}
    \hfill
    \begin{minipage}[t]{0.6\linewidth}
      \centering
      \includegraphics[page=2]{graphics/quadrilateral_locally_minimum_angle}
      \subcaption{Claim~\ref{claim:one-long-diagonal}: $|BC|, |CD| < x$ $\rightarrow$ move $C$ to increase area (left).
        Opposite sides of length $\ge x$ $\rightarrow$ one diagonal $> x$ (right). }%
      \label{fig:claim:one-long-diagonal}
    \end{minipage}
    
    \begin{minipage}[t]{0.6\linewidth}
      \centering
      \includegraphics[page=3]{graphics/quadrilateral_locally_minimum_angle}
      \subcaption{Claim~\ref{claim:diagonal-length-x}: $A \ge 90^\circ$, $|AC| > x$ $\rightarrow$ move $A$ to shorten $AD$ and lengthen $AB$ (left) $\rightarrow$ move $A$ to decrease angle $D$ (right).}%
      \label{fig:claim:diagonal-length-x}
    \end{minipage}
    \hfill
    \begin{minipage}[t]{0.3\linewidth}
      \centering
      \includegraphics[page=4]{graphics/quadrilateral_locally_minimum_angle}
      \subcaption{Claim~\ref{claim:summit-length-x}: $|AB|, |BD| > x$, $B < 90^\circ$ $\rightarrow$ move $B$ to shrink~$C$.}%
      \label{fig:claim:summit-length-x}
    \end{minipage}
    
    \begin{minipage}[t]{0.3\linewidth}
      \centering
      \includegraphics[page=5]{graphics/quadrilateral_locally_minimum_angle}
      \subcaption{Claim~\ref{claim:leg-length-x}: $B < 90^\circ$, $|BC| < x$ $\rightarrow$ move $B$ to lengthen $AB$.}%
      \label{fig:claim:leg-length-x}
    \end{minipage}
    \hfill
    \begin{minipage}[t]{0.6\linewidth}
      \centering
      \includegraphics[page=6]{graphics/quadrilateral_locally_minimum_angle}
      \subcaption{Claim~\ref{claim:base-length-x}: $|CD| < x$, $C > 90^\circ$ $\rightarrow$ move $C$ to shrink angle $C$ (left).
        $|CD| < x$, $C \le 90^\circ$ $\rightarrow$ move $C$ to shrink $C$ and $D$ (right).}%
      \label{fig:claim:base-length-x}
    \end{minipage}
    \caption{Visualization of the different arguments used in Lemma~\ref{lem:quadri1}.
      Most arguments involve moving a corner to shrink the angle sum $C + D$ without violating the length constraints, which shows that the configuration before the movement did not minimize $C + D$.}%
    \label{fig:quadri1}
  \end{figure}

  The following claim does not directly show that one of the segments has the desired length.
  However, it tells us that at least one diagonal has some slack in the sense that slightly moving one of its endpoints does not violate its length constraint.

  \begin{claim}%
    \label{claim:one-long-diagonal}
    At least one diagonal $AC$ or $BD$ is strictly longer than $x$.
  \end{claim}
  \begin{claimproof}
    To prove Claim~\ref{claim:one-long-diagonal}, we first show that there are two opposite sides that have length at least $x$; see Figure~\ref{fig:claim:one-long-diagonal} (left).
    As the summit $AB$ has length at least $x$ by assumption, we need to show that the base $CD$ has length $x$ or both legs $BC$ and $DA$ have length $x$.
    For this, assume that the base $CD$ and one leg, say $BC$, are both shorter than $x$.
    We claim that moving $C$ away from its opposite corner $A$ along the line through $A$ and $C$ shrinks the angle sum $C + D$ without violating any length constraints (if the movement is sufficiently small).
    Concerning the length constraints, the length of the diagonal $AC$ only increases, i.e., $|AC| \ge x$ remains true.
    Additionally, the assumption was that both sides $CD$ and $BC$ are strictly smaller than $x$ and thus moving $C$ by a sufficiently small amount in any direction does not violate $|BC|, |CD| \le x$.
    Concerning the angle sum, note that the area of the quadrilateral increases by the movement, which means that the angle sum $A + B + C + D$ decreases.
    Also note that $A$ remains unchanged and $B$ increases, which means that $C + D$ must decrease.
    Thus, we get a contradiction to $ABCD$ already having minimum angle sum $C + D$.
    It follows that the base $CD$ and the leg $BC$ cannot be both strictly shorter than $x$.
    The argument for the other leg $DA$ is symmetric and thus the base has length $x$ or both legs have length $x$.
    
    It follows that two opposite sides of the quadrilateral both have length at least $x$.
    Using the triangular inequality, this more or less directly implies that not both diagonals can have length $x$.
    If both had length $x$, then their sum would be $2x$.
    Thus, the orange and the blue polygonal lines in Figure~\ref{fig:claim:one-long-diagonal} (right) would have combined length of $2x$ which means one of them has length at most $x$.
    But if the orange (or blue) polyline has length $x$, we get a triangle in which one side has length $\ge x$ and the other two sides together have length $x$.
    This makes the triangular inequality tight, which degenerates the triangle.
  \end{claimproof}

  Now we can start with actually proving that certain segments have the desired length.
  We start with the diagonal $AC$ in case the angle at $A$ is at least $90^\circ$.
  Then we consider the length of the summit $AB$, then the length of the legs $BC$ and $DA$, and finally consider the length of the base $CD$.
 
  \begin{claim}%
    \label{claim:diagonal-length-x}
    If angle $A$ (or $B$) is at least $90^\circ$, then the diagonal $AC$ (or $BD$) has length $x$.
  \end{claim}
  \begin{claimproof}
    Note that the two cases are symmetric, so we only consider the case $A \ge 90^\circ$ and show that the diagonal $AC$ has length $x$.
    Assume for contradiction that $|AC| > x$.
    We show that we can do a series of two movements of $A$ that shrink the angle $D$ without violating the length constraints, which is a contradiction to the assumption that $C + D$ is minimum; see Figure~\ref{fig:claim:diagonal-length-x}.

    The first movement is to move $A$ slightly towards $D$ along the segment $DA$.
    Note that this does not change the angle sum $C + D$.
    However, it shrinks the leg $DA$, which yields $|DA| < x$.
    Moreover, due to the angle of at least $90^\circ$ at $A$, it makes the summit $AB$ longer, which yields $|AB| > x$.
    Also, we can make the movement sufficiently small such that the diagonal $AC$ remains strictly longer than $x$.
    This leaves us in a situation where all three segments $AB$, $AC$, and $AD$ incident to $A$ have slack in the sense that changing the position of $A$ slightly does not violate their length constraints.
    We use this to move $A$ slightly towards $B$ along the side $AB$.
    This clearly shrinks the angle $D$ without changing the angle $C$.
    Thus, the two movements together shrink $C + D$ without violating the length constraints.
  \end{claimproof}

  \begin{claim}%
    \label{claim:summit-length-x}
    The summit $AB$ has length $x$.
  \end{claim}
  \begin{claimproof}
    By Claim~\ref{claim:one-long-diagonal}, one of the diagonals has length greater than $x$.
    We assume without loss of generality that this is true for the diagonal $BD$.
    This also means that the angle $B$ is less than $90^\circ$, as otherwise Claim~\ref{claim:diagonal-length-x} would imply that $BD$ has length $x$.
    Thus, we have $|BD| > x$ and $B < 90^\circ$.

    Now to prove the claim, assume for contradiction that $|AB| > x$.
    Then we move $B$ slightly towards $A$ along the segment $AB$; see Figure~\ref{fig:claim:summit-length-x}.
    We claim that this movement shrinks the angle sum $C + D$ without violating the length constraints, which is a contradiction to the minimality of $C + D$.
    Concerning the angles, the movement clearly shrinks the angle $C$, while the angle $D$ remains unchanged.
    Concerning the length constraints, note that moving $B$ only changes the length of the summit $AB$, of the diagonal $BD$, and of the leg $BC$.
    The summit $AB$ and the diagonal $BD$ are both strictly longer than $x$, so no sufficiently small movement violates their length constraints.
    Additionally, as the angle at $B$ is less than $90^\circ$, moving $B$ towards $A$ shrinks the leg $BC$, which does clearly not violate the constraint $|BC| \le x$.
  \end{claimproof}

  \begin{claim}%
    \label{claim:leg-length-x}
    If the angle $B$ (or $A$) is less than $90^\circ$, the leg $BC$ (or $DA$) has length $x$.
  \end{claim}
  \begin{claimproof}
    We only consider the case where $B < 90^\circ$ and show that $BC$ has length $x$.
    The other case is symmetric.
    Assume for contradiction that $|BC| < x$.
    Then moving $B$ away from $C$ along the line through $BC$ does not change the angles $C$ or $D$; see Figure~\ref{fig:claim:leg-length-x}.
    Concerning the length of segments incident to $B$, note that $BC$ was allowed to get longer due to $|BC| < x$.
    Moreover, as the angle $B$ is less than $90^\circ$, the movement lengthens both the diagonal $BD$ as well as the summit $AB$.
    Thus, the resulting quadrilateral has the same supposedly minimum angle sum $C + D$, but the summit $AB$ is longer than $x$.
    This is a contradiction to Claim~\ref{claim:summit-length-x} (or to put it differently, we can follow up the movement described here with the movement described in the proof of Claim~\ref{claim:summit-length-x} and illustrated in Figure~\ref{fig:claim:summit-length-x} to actually shrink the angle sum).
  \end{claimproof}

  \begin{claim}%
    \label{claim:base-length-x}
    The base $CD$ has length $x$.
  \end{claim}
  \begin{claimproof}
    Again, we assume for contradiction that $|CD| < x$.
    Due to Claim~\ref{claim:one-long-diagonal}, one of the diagonals $AC$ or $BD$ is strictly longer than $x$.
    Assume without loss of generality that $|AC| > x$.
    Now we make a case distinction between $C > 90^\circ$ and $C \le 90^\circ$.
    First, assume $C > 90^\circ$.
    Then we move $C$ away from $D$ along the line through $C$ and $D$; see Figure~\ref{fig:claim:base-length-x} (left).
    This clearly shrinks the angle $C$.
    Additionally, it does not violate any length constraints for the following reason.
    The base $CD$ and the diagonal $AC$ have slack due to the assumptions $|CD| < x$ and $|AC| > x$.
    Moreover, the leg $BC$ gets shorter due to the angle at $C$ being strictly larger than $90^\circ$.

    For the other case that $C \le 90^\circ$, we move $C$ along the cycle through $C$ with center $B$; see Figure~\ref{fig:claim:base-length-x} (right).
    As above, this clearly does not violate any length constraints as $AC$ and $CD$ both have slack and $BC$ remains unchanged due to the movement along a cycle through $B$.
    Concerning the angles $C$ and $D$, this movement clearly shrinks $C$.
    Additionally, as the angle $C$ is at most $90^\circ$, it also shrinks $D$ and thus the sum $C + D$.
  \end{claimproof}

  It now only remains to put things together.
  First consider the case where both angles $A$ and $B$ are less than $90^\circ$.
  Then by Claim~\ref{claim:summit-length-x}, the summit $AB$ has length $x$.
  By Claim~\ref{claim:leg-length-x} both legs $BC$ and $DA$ have length $x$.
  And by Claim~\ref{claim:base-length-x}, $CD$ has length $x$.
  Thus, all sides have length $x$.

  For the other case that the angle $A$ is at least $90^\circ$, Claim~\ref{claim:diagonal-length-x} states that the diagonal $AC$ has length $x$.
  Additionally, by Claim~\ref{claim:summit-length-x}, the summit $AB$ has length $x$.
  As the other diagonal $BD$ cannot also have length $x$ by Claim~\ref{claim:one-long-diagonal}, the angle at $B$ must be less than $90^\circ$ (contraposition of Claim~\ref{claim:diagonal-length-x}).
  Thus, we can apply Claim~\ref{claim:leg-length-x} to obtain that the leg $BC$ has length $x$.
  This lets us apply Claim~\ref{claim:base-length-x} stating that the base $CD$ has length $x$.
  Thus, overall, we have that all sides except $DA$ have length $x$ and additionally the diagonal $AC$ has length $x$.
  This concludes the proof.
\end{proof}

According to~\cref{lem:quadri1} a quadrilateral with minimum angle sum $C + D$ is in one of two configurations.
In the following we provide a lower bound on $C + D$ for both configurations.
In the first configuration all sides have length $x$; see~\cref{fig:quadri_case1}.
For the second configuration three sides and one diagonal have length $x$ and the remaining side has length at most $x$; see~\cref{fig:quadri_case2}.

\begin{lemma}\label{lem:quadri2}
  Let $ABCD$ be a quadrilateral with minimum angle sum $C + D$ such that all sides have length $x$.
  Then
  \begin{equation*}
	C + D \ge 4\asin\left(\frac{\cos(\pi/4)}{\cosh(x/2)}\right).
  \end{equation*}
\end{lemma} 

\begin{proof}
	Note that the angle sum $C + D$ is equal to the angle sum $A+B$.
	This can be seen by splitting the quadrilateral at the diagonal into two congruent isosceles triangles.
	Consequently, the sum $C+D$ is minimized, when the sum of all internal angles is minimized and the area of the quadrilateral is maximized.
	By \cref{lem:kgon}, the area is maximized when the quadrilateral is regular with angle $\alpha=2\asin\left(\frac{\cos(\pi/4)}{\cosh(x/2)}\right)$ at each vertex and $C+D\geq 2\alpha$.
\end{proof}

\begin{lemma}%
  \label{lem:quadri3}
  Let $ABCD$ be a quadrilateral with $|AB| = |BC| = |CD| = |AC| = x$ and $|DA| \le x$.
  Let $f(D) = C(D) + D$ be the sum of angles at $C$ and $D$ as a function of angle $D$.
  \begin{itemize}
	\item If $\cosh(x) \le 2$, then the function $f$ is monotone decreasing in $D$. In this case 
	\begin{equation*}
			f(D) \geq \frac{\pi}{2} + \acos\left(\frac{\tanh(x/2)}{\tanh(x)}\right).
	\end{equation*}
	\item If $\cosh(x) > 2$, then $f$ attains its unique minimum at $D'=\asin\left(\sqrt{\frac{2\cosh(x)-1}{\sinh^2(x)}} \right)$.
		  In this case
	\begin{equation*}
			f(D) \geq D' + 2\acot(\cosh(x)\tan(D')) + \acos\left(\frac{\tanh(x/2)}{\tanh(x)}\right).
	\end{equation*}
  \end{itemize}
\end{lemma}
\begin{proof}
  Split the quadrilateral at the diagonal $AC$.
  This yields two triangles $ABC$ and $ACD$, see~\cref{fig:quadri_case2}.
  Denote by $\gamma_1$ the angle at $C$ in triangle $ABC$ and by $\gamma_2$ the angle at $C$ in triangle $ACD$, i.e., $C = \gamma_1 + \gamma_2$.
  The triangle $ABC$ is equilateral and thus $\gamma_1$ is uniquely determined by side length $x$ as $\gamma_1 = \acos\left(\frac{\tanh(x/2)}{\tanh(x)}\right)$.

  Consider the isosceles triangle $ACD$ with sides $|AC| = |CD| = x$ and note that $D < \pi/2$.
  Splitting the triangle along the perpendicular bisector of $AD$ and using right-triangle formulas we obtain $\cot(\frac{\gamma_2}{2}) = \cosh(x)\tan(D)$ and hence $\gamma_2 = 2\acot(\cosh(x)\tan(D))$.
  We calculate the derivative of $f(D)=D + \gamma_2(D) + \gamma_1$ using that $\frac{\partial}{\partial x}\acot(c\tan(x)) = -c/(c^2\sin^2(x)+\cos^2(x))$ and the identities $\cos^2(x) + \sin^2(x) = 1$ and $\sinh^2(x)+1=\cosh^2(x)$:
  \begin{align*}
	\frac{\partial}{\partial D}f(D) &= 1 - \frac{2\cosh(x)}{\cos^2(D) + \cosh^2(x)\sin^2(D)} \\
	&= 1 - \frac{2\cosh(x)}{1 + (\cosh^2(x)-1)\sin^2(D)} \\
	&= 1 - \frac{2\cosh(x)}{1+\sin^2(D)\sinh^2(x)}
  \end{align*}
  For $\cosh(x) \le 2$ and using that $\sin^2(D) < 1$ we have: 
  \begin{align*}
	\frac{\partial}{\partial D}f(D) < 1 - \frac{2\cosh(x)}{1+\sinh^2(x)} = 1 - \frac{2\cosh(x)}{\cosh^2(x)} = 1 - \frac{2}{\cosh(x)} \leq 0
  \end{align*}
  The lower bound follows by taking the limit.
  As $D$ approaches $\pi/2$, angle $\gamma_2$ approaches $0$ and $\gamma_1$ stays constant, which yields the lower bound. 
  For $\cosh(x) > 2$ the angle sum $f(D)$ attains its minimum at $\frac{\partial}{\partial D}f(D) = 0$ which is the case if and only if $D=\asin\left(\sqrt{\frac{2\cosh(x)-1}{\sinh^2(x)}} \right)$.
\end{proof} 

\subsection{Putting Things Together}%
\label{sec:putt-things-togeth}
We are now ready to prove that there exist radii $s_1 < s < s_2$ such that the graph $\tilde C$ in~\cref{fig:c_k_l} has sphericity $2$ for $s_1$ and $s_2$ but higher sphericity for $s$.

\begin{theorem}\label{thm:jump_example}
	There are constants $s_1 < s_\ell < s_u < s_2$ such that $\HSph{\tilde C}{s}=3$ for all $s \in [s_\ell, s_u]$ and $\HSph{\tilde C}{s_1}=\HSph{\tilde C}{s_2}=2$.
\end{theorem}

\begin{proof}
	We use a force-directed method to compute $2$-dimensional HUBG representations of $\tilde C$ for different values of $s$.
	For $s_1 = 0.7635$ and $s_2 = 0.7795$ the graph $\tilde C$ is a $2$-dimensional HUBG, as shown in \cref{fig:c_k_l_sol1,fig:c_k_l_sol2}.
  To see that $\tilde C$ has no $2$-dimensional HUBG representation for $s \in [s_\ell, s_u]$, note that $\tilde C$ contains no triangles.
  By \cref{lem:planar_drawing} every drawing induced by a $2$-dimensional HUBG representation of $\tilde C$ is planar.
	As a result, a path must lie entirely inside or outside the cycle.
	In the remainder of this proof we show that for radius $s\geq s_\ell = 0.765$ it is not possible to place an independent vertex inside the cycle of $\tilde C$, 
	and that for $s\leq s_u=0.770$ it is not possible to place all paths outside the cycle.
	This gives us the desired interval.

	We first show that for $s\geq s_\ell$ it is not possible to place an independent vertex inside a $k=8$ cycle.
	Assume that we have a HUBG representation with radius $s$ containing an independent vertex $p'$ inside.
	Two consecutive cycle vertices form a triangle together with $p'$ with angle $\alpha$ at $p'$.
	These $k$ angles must be enough to cover the full $2\pi$ angle around $p'$, i.e., the largest angle must be at least $2\pi/k$.
	Consider two consecutive cycle vertices.
	Because both cycle vertices are unconnected with $p'$ but connected with each other, 
	two sides of this triangle are larger than $2s$ and one side is smaller than $2s$.
	Using \cref{lem:extrem_triangle2}, we obtain that $\alpha \leq 2\asin\left(\sinh(s)/\sinh(2s)\right)$.
	For $s_\ell$ we obtain $\alpha \approx 0.7850 < 2\pi/8\approx 0.7854$.
	As $\alpha$ is decreasing with growing $s$, it is not possible to place an independent vertex inside the cycle for $s\geq s_\ell$.
	
	In the following, we show that for $s\leq s_u$ it is not possible to place all $p$ paths outside the cycle.
	Consider the angles between consecutive neighbors of a cycle vertex.
	We classify these angles into~$3$ different types and give lower bounds for each, see~\cref{fig:angles}.
	Type $1$ angles are between two path neighbors, type $2$ angles between a path neighbor and a cycle neighbor, and type $3$ angles are between two cycle neighbors.
	In total, there are $p-k$ angles of type $1$, $2k$ angles of type $2$ and $k$ angles of type $3$.
	We prove a lower bound for the sum of these angles that increases with decreasing $s$.	
	For a valid HUBG representation this sum must not exceed $2\pi k$, the angle sum around all cycle vertices.
 
	\subparagraph*{Type 1.} Consider two consecutive path neighbors $p_1, p_2$ of a cycle vertex $c_1$.
	They form a triangle, where the two shorter sides have length $\leq 2s$ and the long side has length $> 2s$.
	Using \cref{lem:extrem_triangle1} we obtain that the angle at $c_1$ is at least $A=2\asin\left(\sinh(s)/\sinh(2s)\right)$.

	\subparagraph*{Type 2.} We consider pairs of type $2$ angles, see angle $\beta$ (blue) in \cref{fig:angles}.
	Two neighboring cycle vertices $c_1, c_2$ and their respective path vertices $p_1, p_2$ form a quadrilateral.
	We are interested in a lower bound $B$ for the sum of the two angles at $c_1$ and $c_2$.
	The side length between $p_1$ and $p_2$ and the diagonals are at least $2s$, while the other three sides are at most $2s$.
	By \cref{lem:quadri1} we know that we can apply \cref{lem:quadri2} or \cref{lem:quadri3} to obtain a lower bound. 
	Let $f_1(2s)$ be the lower bound obtained from \cref{lem:quadri2} and $f_2(2s)$ be the one obtained from \cref{lem:quadri3}.
	Then we have $B = \min(f_1(2s), f_2(2s))$.
	
	\subparagraph*{Type 3.} The sum of all type $3$ angles is exactly the internal angle sum of the $k$-gon induced by the cycle vertices.
	All sides of this $k$-gon are $\leq 2s$.
	The area $A$ of the $k$-gon relates to the sum of internal angles $S$ via the identity $A = (k-2)\pi - S$.
	It follows, that the angles are minimized when the area is maximized.
	Using \cref{lem:kgon}, we see that this is the case, when the $k$-gon is regular with internal angle sum $C=2k\asin\left( \frac{\cos(\pi/k)}{\cosh(s)} \right)$.

	\subparagraph*{Sum of all Angles.} If the sum of all type $1,2$ and $3$ angles exceed $2\pi k$ it is not possible to place all paths outside the cycle.
	Putting everything together we obtain that the sum of all angles is $S = (p-k)A + kB + C$.
	Moreover, $S$ is decreasing with increasing $s$.
	For $s_u=0.770$, $k=8$ and $p=33$ we can calculate that $50.290 \approx S > 2\pi k \approx 50.265$.
	Together with the previous result, we obtain that for $s \in [0.765, 0.770]$ the hyperbolic sphericity for $\tilde C$ is at least $3$ and using \cref{thm:large-radius} exactly $3$.
\end{proof}

Both drawings in \cref{fig:c_k_l_sol1,fig:c_k_l_sol2} were computed using a force-directed method.
We note, that the drawing in \cref{fig:c_k_l_sol2} can also be realized using a larger radius, as this only increases the available space for the paths.
The drawing in \cref{fig:c_k_l_sol1} however, requires a more careful consideration.
To obtain a non-computer-aided drawing, one can start by placing the graph vertices on the vertices of a regular $\{3,8\}$-tiling of the hyperbolic plane.
Both vertices of one path are placed in the center, surrounded by $8$ cycle vertices, which are surrounded by the remaining $32$ path vertices.
This configuration is not a valid HUBG representation yet, as the balls of some unconnected vertices are touching.
However, it does not minimize the type $3$ angles.
By rotating the outer layer of path vertices, the type $3$ angles become smaller and allow for enough slack to decrease the radius slightly 
and obtain a valid HUBG representation.

\section{Open Problems}\label{sec:open_problems}

We conclude by highlighting several open problems that we believe are worth exploring.

It remains unclear whether the hyperbolic sphericity can grow linearly with the size of the graph. 
For Euclidean or dot-product representations, 
it is known that the complete bipartite graph $K_{n,n}$ requires at least $n$ dimensions. 
We ask whether a similar requirement holds in the hyperbolic case. 
Unfortunately, the proof for the Euclidean setting cannot be easily transferred, although a logarithmic lower bound can be obtained.
Since non-uniform ball graphs (i.e., ball graphs where the radii of the balls can vary) are a generalization of hyperbolic unit ball graphs, 
a lower bound on their dimension directly applies to hyperbolic unit ball graphs as well. 
However, to the best of our knowledge, the best known lower bound for non-uniform ball graphs only grows logarithmically as well.

In~\cref{thm:large-radius} we demonstrated that, for a sufficiently large radius, 
the hyperbolic sphericity can be at most~$1$ larger than its minimum.
Given a graph, however, it is not immediately clear when a radius is \emph{large enough}.
It would be interesting to determine whether a lower bound for the hyperbolic radius $s^\ast$ exists (dependent only on the graph size), such that the hyperbolic sphericity with radius $s^\ast$ is at most $1$~larger than the minimum hyperbolic sphericity; a question already stated in~\cite{BlasiusDHM26}. 
We suspect that stars require the largest lower bound.
If this is the case, $s^\ast$ would grow logarithmically with the graph size.

Another open question is how large the difference between Euclidean and hyperbolic sphericity can become asymptotically.
Here, too, stars are a canonical candidate. 
In this case, an EUBG representation requires at most logarithmically more dimensions than a HUBG representation.
It is not obvious, whether this is the largest possible gap between Euclidean and hyperbolic sphericity.

\begin{figure}
	\centering 
	\includegraphics{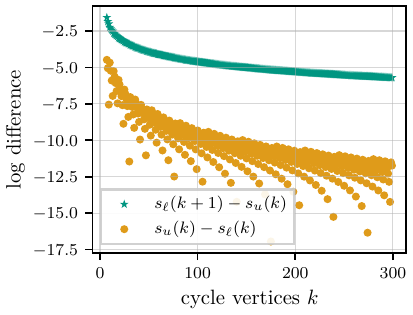}
	\caption{Difference between lower and upper bounds of \cref{thm:jump_example}.
	The $x$-axis indicates the number of cycle vertices in $\tilde C(k)$.
	The $y$-axis shows the logarithmic difference between bounds.
	Circles (orange) indicate the size of the interval, stars (green) indicate the distance between the intervals.
	}\label{fig:bound_difference}
\end{figure}

Finally, we conjecture that it is possible to construct a family of graphs where the hyperbolic sphericity oscillates arbitrarily often between two values as the radius increases. 
To achieve this, we can take the disjoint union of different variations of the graph $\tilde C$ from \cref{thm:jump_example}.
By choosing a varying amount of cycle vertices, each with an appropriate number of paths,
we obtain a distinct graph $\tilde C(k)$ for each $k$, yielding lower and upper bounds $s_\ell(k)$ and $s_u(k)$ for impossible $2$-dimensional HUBG representations.
In \cref{fig:bound_difference} we compute $s_\ell(k)$ and $s_u(k)$ for $k\in [7, 300]$ and plot their difference.
The size of these intervals $s_u(k) - s_\ell(k)$ is always positive, indicating that for all $k$ there is an interval of radii where the hyperbolic sphericity of $\tilde C(k)$ is at least~$3$.
Additionally, the distance between intervals $s_\ell(k+1) - s_u(k)$ is also always positive, showing that they do not overlap.
It is also possible to consider cases, where more than one path is placed inside the cycle to yield further configurations.
However, constructing a single oscillating graph requires additional work as we must be careful to ensure that the invalid regions (regions for which no $2$-dimensional HUBG representation exists) of one graph do not overlap too much with the valid regions of the other graphs.



\bibliography{hyperbolic-sphericity-paper}

@book{Ratcliffe,
    title = {Foundations of Hyperbolic Manifolds},
    author = {Ratcliffe, J.},
    isbn = {978-3-030-31597-9},
    series = {Graduate Texts in Mathematics},
    year = {2019},
    doi = {10.1007/978-3-030-31597-9},
    edition = {3rd},
}

@book{maxima-and-minima,
  title={Maxima and minima without calculus},
  author={Niven, Ivan},
  volume={6},
  year={1981},
  publisher={American Mathematical Soc.}
}

@book{convex-optimization,
  title     = {Convex Optimization},
  author    = {Stephen P. Boyd and Lieven Vandenberghe},
  year      = {2004},
  url       = {https://web.stanford.edu/~boyd/cvxbook/},
  publisher={Cambridge University Press},
  doi       = {10.1017/CBO9780511804441}
}

@article{Mae-Main,
  title     = {Space graphs and sphericity},
  author    = {Hiroshi Maehara},
  pages     = {55--64},
  year      = {1984},
  journal   = {Discrete Applied Mathematics},
  volume    = {7},
  number    = {1},
  doi       = {10.1016/0166-218X(84)90113-6}
}

@inproceedings{BlasiusDHM26,
  title     = {Product Structure and Treewidth of Hyperbolic Uniform Disk Graphs},
  author    = {Thomas Bläsius and Emil Dohse and Deborah Haun and Laura Merker},
  pages     = {18:1--18:17},
  year      = {2026},
  booktitle = {International Symposium on Computational Geometry (SoCG)},
  doi       = {10.4230/LIPIcs.SoCG.2026.18}
}

@article{Mae-Knn,
  title     = {Sphericity exceeds cubicity for almost all complete bipartite graphs},
  author    = {Hiroshi Maehara},
  journal   = {Journal of Combinatorial Theory, Series B},
  pages     = {231--235},
  year      = {1986},
  volume    = {40},
  number    = {2},
  doi       = {10.1016/0095-8956(86)90081-X}
}

@article{Mae-Frankl-JL-Lemma,
  title     = {The Johnson-Lindenstrauss lemma and the sphericity of some graphs},
  author    = {Peter Frankl and Hiroshi Maehara},
  journal   = {Journal of Combinatorial Theory, Series B},
  pages     = {355--362},
  year      = {1988},
  volume    = {44},
  number    = {3},
  doi       = {10.1016/0095-8956(88)90043-3}
}

@article{HeinzUnitDiskRecocNPHard,
  title     = {Unit disk graph recognition is NP-hard},
  author    = {Heinz Breu and David G. Kirkpatrick},
  pages     = {3--24},
  year      = {1998},
  journal   = {Computational Geometry},
  volume    = {9},
  number    = {1-2},
  doi       = {10.1016/S0925-7721(97)00014-X}
}

@inproceedings{SchaeferDiskRecocERHard,
  title     = {Complexity of Some Geometric and Topological Problems},
  author    = {Marcus Schaefer},
  pages     = {334--344},
  year      = {2009},
  booktitle = {International Symposium on Graph Drawing and Network Visualization (GD)},
  doi       = {10.1007/978-3-642-11805-0_32}
}

@inproceedings{blaesiusStrongyHyperbolic,
  title     = {Strongly Hyperbolic Unit Disk Graphs},
  author    = {Thomas Bläsius and Tobias Friedrich and Maximilian Katzmann and Daniel Stephan},
  pages     = {13:1--13:17},
  year      = {2023},
  booktitle = {Symposium on Theoretical Aspects of Computer Science (STACS)},
  doi       = {10.4230/LIPIcs.STACS.2023.13}
}

@article{bachEdgeBundling,
  title     = {Towards Unambiguous Edge Bundling: Investigating Confluent Drawings for Network Visualization},
  author    = {Benjamin Bach and Nathalie Henry Riche and Christophe Hurter and Kim Marriott and Tim Dwyer},
  pages     = {541--550},
  year      = {2017},
  journal   = {IEEE Transactions on Visualization and Computer Graphics (TVCG)},
  volume    = {23},
  number    = {1},
  doi       = {10.1109/TVCG.2016.2598958}
}

@article{voitalovScaleFree,
    title = {Scale-free networks well done},
    author = {Voitalov, Ivan and van der Hoorn, Pim and van der Hofstad, Remco
              and Krioukov, Dmitri},
    journal = {Physical Review Research},
    volume = {1},
    number = {3},
    pages = {033034},
    year = {2019},
    doi = {10.1103/PhysRevResearch.1.033034},
}

@article{krioukovComplexNetworks,
  title = {Hyperbolic geometry of complex networks},
  author = {Krioukov, Dmitri and Papadopoulos, Fragkiskos and Kitsak, Maksim and Vahdat, Amin and Bogu\~n\'a, Mari\'an},
  journal = {Physical Review E},
  volume = {82},
  number = {3},
  pages = {036106},
  year = {2010},
  doi = {10.1103/PhysRevE.82.036106},
}

@inproceedings{SakarLowDistortionTrees,
  title     = {Low Distortion Delaunay Embedding of Trees in Hyperbolic Plane},
  author    = {Rik Sarkar},
  pages     = {355--366},
  year      = {2011},
  booktitle = {International Symposium on Graph Drawing and Network Visualization (GD)},
  doi       = {10.1007/978-3-642-25878-7_34}
}

@article{robertsBoxicityCubicity,
    title = {On the boxicity and cubicity of a graph},
    author = {Roberts, Fred S},
    journal = {Recent progress in combinatorics},
    volume = {1},
    number = {1},
    pages = {301--310},
    year = {1969},
}

@inproceedings{sandorHyperbolicBall,
  title     = {Hyperbolic intersection graphs and (quasi)-polynomial time},
  author    = {Sándor Kisfaludi-Bak},
  pages     = {1621--1638},
  year      = {2020},
  booktitle = {ACM-SIAM Symposium on Discrete Algorithms (SODA)},
  doi       = {10.1137/1.9781611975994.100}
}

@inproceedings{blaesiusStructureAndIndependence,
  title     = {Structure and Independence in Hyperbolic Uniform Disk Graphs},
  author    = {Thomas Bläsius and Jean-Pierre von der Heydt and Sándor Kisfaludi-Bak and Marcus Wilhelm and Geert van Wordragen},
  pages     = {21:1--21:16},
  year      = {2025},
  booktitle = {International Symposium on Computational Geometry (SoCG)},
  doi       = {10.4230/LIPIcs.SoCG.2025.21}
}

@inproceedings{eppsteinLimitsHyperbolicGraphDrawing,
  title     = {Limitations on Realistic Hyperbolic Graph Drawing},
  author    = {David Eppstein},
  pages     = {343--357},
  year      = {2021},
  booktitle = {International Symposium on Graph Drawing and Network Visualization (GD)},
  doi       = {10.1007/978-3-030-92931-2_25}
}

@inproceedings{gugelmannHyperbolicDegreeSequence,
  title     = {Random Hyperbolic Graphs: Degree Sequence and Clustering - (Extended Abstract)},
  author    = {Luca Gugelmann and Konstantinos Panagiotou and Ueli Peter},
  pages     = {573--585},
  year      = {2012},
  booktitle = {International Colloquium on Automata, Languages and Programming (ICALP)},
  doi       = {10.1007/978-3-642-31585-5_51}
}

@inproceedings{BlaesiusHyperbolicSeparatorsAndTreewidth,
  title     = {Hyperbolic Random Graphs: Separators and Treewidth},
  author    = {Thomas Bläsius and Tobias Friedrich and Anton Krohmer},
  pages     = {15:1--15:16},
  year      = {2016},
  booktitle = {European Symposium on Algorithms (ESA)},
  doi       = {10.4230/LIPIcs.ESA.2016.15}
}

@article{FriedrichDimeterHyperbolicRandomGraphs,
  title     = {On the Diameter of Hyperbolic Random Graphs},
  author    = {Tobias Friedrich and Anton Krohmer},
  pages     = {1314--1334},
  year      = {2018},
  journal   = {SIAM Journal on Discrete Mathematics (SIDMA)},
  volume    = {32},
  number    = {2},
  doi       = {10.1137/17M1123961}
}

@article{BlaesiusCliquesHyperbolicRandomGraphs,
  title     = {Cliques in Hyperbolic Random Graphs},
  author    = {Thomas Bläsius and Tobias Friedrich and Anton Krohmer},
  pages     = {2324--2344},
  year      = {2018},
  journal   = {Algorithmica},
  volume    = {80},
  number    = {8},
  doi       = {10.1007/s00453-017-0323-3}
}

@article{blaesiusVertexCover,
  title     = {Solving Vertex Cover in Polynomial Time on Hyperbolic Random Graphs},
  author    = {Thomas Bläsius and Philipp Fischbeck and Tobias Friedrich and Maximilian Katzmann},
  pages     = {28--51},
  year      = {2023},
  journal   = {Theory of Computing Systems},
  volume    = {67},
  number    = {1},
  doi       = {10.1007/s00224-021-10062-9}
}

@inproceedings{KiwiBoundDiameter,
  title     = {A Bound for the Diameter of Random Hyperbolic Graphs},
  author    = {Marcos A. Kiwi and Dieter Mitsche},
  pages     = {26--39},
  year      = {2015},
  booktitle = {Workshop on Analytic Algorithmics and Combinatorics (ANALCO)},
  doi       = {10.1137/1.9781611973761.3}
}

@inproceedings{BaguleyCliqueNumberDegeneracy,
  title     = {Hyperbolic Random Graphs: Clique Number and Degeneracy with Implications for Colouring},
  author    = {Samuel Baguley and Yannic Maus and Janosch Ruff and George Skretas},
  pages     = {13:1--13:20},
  year      = {2025},
  booktitle = {Symposium on Theoretical Aspects of Computer Science (STACS)},
  doi       = {10.4230/LIPIcs.STACS.2025.13}
}

@inproceedings{EnujinComputingMaximumCliques,
  title     = {Algorithms for Computing Maximum Cliques in Hyperbolic Random Graphs},
  author    = {Eunjin Oh and Seunghyeok Oh},
  pages     = {85:1--85:15},
  year      = {2023},
  booktitle = {European Symposium on Algorithms (ESA)},
  doi       = {10.4230/LIPIcs.ESA.2023.85}
}

@article{blaesiusVertexCoverApproximation,
  title     = {Efficiently Approximating Vertex Cover on Scale-Free Networks with Underlying Hyperbolic Geometry},
  author    = {Thomas Bläsius and Tobias Friedrich and Maximilian Katzmann},
  pages     = {3487--3520},
  year      = {2023},
  journal   = {Algorithmica},
  volume    = {85},
  number    = {12},
  doi       = {10.1007/s00453-023-01143-x}
}

@article{blaesiusShortestPathHyperbolic,
  title     = {Efficient Shortest Paths in Scale-Free Networks with Underlying Hyperbolic Geometry},
  author    = {Thomas Bläsius and Cedric Freiberger and Tobias Friedrich and Maximilian Katzmann and Felix Montenegro-Retana and Marianne Thieffry},
  pages     = {19:1--19:32},
  year      = {2022},
  journal   = {ACM Transactions on Algorithms (TALG)},
  volume    = {18},
  number    = {2},
  doi       = {10.1145/3516483}
}

@article{jacomyForceAtlas2,
    doi = {10.1371/journal.pone.0098679},
    author = {Jacomy, Mathieu AND Venturini, Tommaso AND Heymann, Sebastien AND
              Bastian, Mathieu},
    journal = {PLOS ONE},
    title = {ForceAtlas2, a Continuous Graph Layout Algorithm for Handy Network
             Visualization Designed for the Gephi Software},
    year = {2014},
    volume = {9},
    pages = {1--12},
    number = {6},
}

@article{muellerDiameterKPKVB,
    title = {The diameter of KPKVB random graphs},
    volume = {51},
    doi = {10.1017/apr.2019.23},
    number = {2},
    journal = {Advances in Applied Probability},
    author = {Müller, Tobias and Staps, Merlijn},
    year = {2019},
    pages = {358--377},
}

@article{michelLargestComponentHyperbolic,
  title     = {On the Largest Component of a Hyperbolic Model of Complex Networks},
  author    = {Michel Bode and Nikolaos Fountoulakis and Tobias Müller},
  pages     = {3},
  year      = {2015},
  journal   = {The Electronic Journal of Combinatorics},
  volume    = {22},
  number    = {3},
  doi       = {10.37236/4958}
}

@article{marcosSecondLargestComponent,
  title     = {On the Second Largest Component of Random Hyperbolic Graphs},
  author    = {Marcos A. Kiwi and Dieter Mitsche},
  pages     = {2200--2217},
  year      = {2019},
  journal   = {SIAM Journal on Discrete Mathematics (SIDMA)},
  volume    = {33},
  number    = {4},
  doi       = {10.1137/18M121201X}
}

@article{ReitermanKnnDotProductDim,
  title     = {Geometrical embeddings of graphs},
  author    = {Jan Reiterman and Vojtech Rödl and Edita Sinajová},
  pages     = {291--319},
  year      = {1989},
  journal   = {Discrete Mathematics},
  volume    = {74},
  number    = {3},
  doi       = {10.1016/0012-365X(89)90142-8}
}

@article{lampingHyperbolicBrowser,
  title     = {The Hyperbolic Browser: A Focus + Context Technique for Visualizing Large Hierarchies},
  author    = {John Lamping and Ramana Rao},
  pages     = {33--55},
  year      = {1996},
  journal   = {Journal of Computer Languages (COLA)},
  volume    = {7},
  number    = {1},
  doi       = {10.1006/jvlc.1996.0003}
}

@article{MunzerHyperbolic3DDrawing1,
  title     = {Exploring Large Graphs in 3D Hyperbolic Space},
  author    = {Tamara Munzner},
  pages     = {18--23},
  year      = {1998},
  journal   = {IEEE Computer Graphics and Applications},
  volume    = {18},
  number    = {4},
  doi       = {10.1109/38.689657}
}

@inproceedings{MunzerHyperbolic3DDrawing2,
  title     = {H3: laying out large directed graphs in 3D hyperbolic space},
  author    = {Tamara Munzner},
  pages     = {2--10},
  year      = {1997},
  booktitle = {IEEE Information Visualization Conference (InfoVis)},
  doi       = {10.1109/INFVIS.1997.636718}
}

@inproceedings{MunzerHyperbolic3DDrawing3,
  title     = {Visualizing the Structure of the World Wide Web in 3D Hyperbolic Space},
  author    = {Tamara Munzner and Paul Burchard},
  pages     = {33--38},
  year      = {1995},
  booktitle = {International Conference on 3D Technologies for the World Wide Web (Web3D)},
  doi       = {10.1145/217306.217311}
}

@inproceedings{MunzerHyperbolic3DDrawing4,
  title     = {Drawing Large Graphs with H3Viewer and Site Manager},
  author    = {Tamara Munzner},
  pages     = {384--393},
  year      = {1998},
  booktitle = {International Symposium on Graph Drawing and Network Visualization (GD)},
  doi       = {10.1007/3-540-37623-2_30}
}

@article{blaesiusEmbeddingHyperbolicPlane,
  title     = {Efficient Embedding of Scale-Free Graphs in the Hyperbolic Plane},
  author    = {Thomas Bläsius and Tobias Friedrich and Anton Krohmer and Sören Laue},
  pages     = {920--933},
  year      = {2018},
  journal   = {IEEE Transactions on Networking (TON)},
  volume    = {26},
  number    = {2},
  doi       = {10.1109/TNET.2018.2810186}
}

@article{stephenNonEuclideanEmbedder,
  title     = {Non-Euclidean Spring Embedders},
  author    = {Stephen G. Kobourov and Kevin Wampler},
  pages     = {757--767},
  year      = {2005},
  journal   = {IEEE Transactions on Visualization and Computer Graphics (TVCG)},
  volume    = {11},
  number    = {6},
  doi       = {10.1109/TVCG.2005.103}
}

@inproceedings{blaesiusForceDirected,
  title     = {Force-Directed Embedding of Scale-Free Networks in the Hyperbolic Plane},
  author    = {Thomas Bläsius and Tobias Friedrich and Maximilian Katzmann},
  pages     = {22:1--22:18},
  year      = {2021},
  booktitle = {Symposium on Experimental and Efficient Algorithms (SEA)},
  doi       = {10.4230/LIPIcs.SEA.2021.22}
}

@article{fragkiskosNetworkMapping,
  title     = {Network Mapping by Replaying Hyperbolic Growth},
  author    = {Fragkiskos Papadopoulos and Constantinos Psomas and Dmitri V. Krioukov},
  pages     = {198--211},
  year      = {2015},
  journal   = {IEEE Transactions on Networking (TON)},
  volume    = {23},
  number    = {1},
  doi       = {10.1109/TNET.2013.2294052}
}

@article{fragkiskosNetowrInferenceCommonNeighbors,
  title = {Network geometry inference using common neighbors},
  author = {Papadopoulos, Fragkiskos and Aldecoa, Rodrigo and Krioukov, Dmitri},
  journal = {Physical Review E},
  volume = {92},
  number = {2},
  pages = {022807},
  year = {2015},
  doi = {10.1103/PhysRevE.92.022807},
}

@article{blaesiusHyperbolicGreedyRouting,
  title     = {Hyperbolic Embeddings for Near-Optimal Greedy Routing},
  author    = {Thomas Bläsius and Tobias Friedrich and Maximilian Katzmann and Anton Krohmer},
  pages     = {1--18},
  year      = {2020},
  journal   = {ACM Journal of Experimental Algorithmics (JEA)},
  volume    = {25},
  doi       = {10.1145/3381751}
}

@article{eppsteinSuccinctGreedyRouting,
  title     = {Succinct Greedy Geometric Routing Using Hyperbolic Geometry},
  author    = {David Eppstein and Michael T. Goodrich},
  pages     = {1571--1580},
  year      = {2011},
  journal   = {IEEE Transactions on Computers},
  volume    = {60},
  number    = {11},
  doi       = {10.1109/TC.2010.257}
}

@inproceedings{kleinbergGeographicRouting,
  title     = {Geographic Routing Using Hyperbolic Space},
  author    = {Robert Kleinberg},
  pages     = {1902--1909},
  year      = {2007},
  booktitle = {IEEE Conference on Computer Communications (INFOCOM)},
  doi       = {10.1109/INFCOM.2007.221}
}

@inproceedings{node2vec,
  title     = {node2vec: Scalable Feature Learning for Networks},
  author    = {Aditya Grover and Jure Leskovec},
  pages     = {855--864},
  year      = {2016},
  booktitle = {ACM SIGKDD Conference on Knowledge Discovery and Data Mining (KDD)},
  doi       = {10.1145/2939672.2939754}
}

@inproceedings{deepwalk,
  title     = {DeepWalk: online learning of social representations},
  author    = {Bryan Perozzi and Rami Al-Rfou and Steven Skiena},
  pages     = {701--710},
  year      = {2014},
  booktitle = {ACM SIGKDD Conference on Knowledge Discovery and Data Mining (KDD)},
  doi       = {10.1145/2623330.2623732}
}

@inproceedings{verse,
  title     = {VERSE: Versatile Graph Embeddings from Similarity Measures},
  author    = {Anton Tsitsulin and Davide Mottin and Panagiotis Karras and Emmanuel Müller},
  pages     = {539--548},
  year      = {2018},
  booktitle = {The Web Conference (WWW)},
  doi       = {10.1145/3178876.3186120}
}

@book{battistaGrapDrawing,
  title     = {Graph Drawing: Algorithms for the Visualization of Graphs},
  author    = {Giuseppe Di Battista and Peter Eades and Roberto Tamassia and Ioannis G. Tollis},
  publisher = {Prentice-Hall},
  year      = {1999},
}

@proceedings{kaufmannDrawingGraphs,
  title     = {Drawing Graphs, Methods and Models},
  editor    = {Kaufmann, Michael and Wagner, Dorothea},
  year      = {2001},
  doi       = {10.1007/3-540-44969-8}
}

@article{JohnsonAndLindenstrauss,
	author    = {Johnson, W.B. and Lindenstrauss, J.},
	title     = {Extensions of Lipschitz mappings into a Hilbert space},
	year      = {1984},
	journal   = {Contemporary Mathematics},
	volume    = {26},
	pages     = {189--206},
}

\end{document}